%% file: acl_latex.tex
\documentclass[11pt]{article}

\usepackage[final]{acl}

\usepackage{times}
\usepackage{latexsym}

\usepackage[T1]{fontenc}
\usepackage[utf8]{inputenc}

\usepackage{microtype}

\usepackage{inconsolata}

\usepackage{graphicx}

\usepackage{amsmath}
\usepackage{amsfonts}
\usepackage{booktabs}
\usepackage{multirow}

\usepackage{xcolor}  % 用于控制颜色
\usepackage{amsmath} % 用于数学符号
\usepackage{subcaption}
\usepackage{enumitem}
\usepackage[ruled,linesnumbered]{algorithm2e}

\newcommand{\rdown}[1]{\textcolor{red}{\ensuremath{\downarrow_{#1}}}}

\title{Evo-Attacker: Memory-Augmented Reinforcement Learning for Long-Horizon Tool Attacks on LLM-MAS}

\author{
  Bingyu Yan\textsuperscript{1},
  Xiaoming Zhang\textsuperscript{1}\thanks{Xiaoming Zhang is the corresponding author},
  Jinyu Hou\textsuperscript{2},
  Chaozhuo Li\textsuperscript{2},
  Ziyi Zhou\textsuperscript{1},
  \\
  \textbf{Yiming Hei\textsuperscript{3}},
  \textbf{Litian Zhang\textsuperscript{2}}
\\
  \textsuperscript{1}Beihang University \\
  \textsuperscript{2}Beijing University of Posts and Telecommunications \\
  \textsuperscript{3}China Academy of Information and Communications Technology
}

\begin{document}
\maketitle

\begin{abstract}
While Large Language Model-based Multi-Agent Systems (LLM-MAS) demonstrate remarkable capabilities in solving complex tasks by orchestrating specialized agents and external tools, the implicit trust in tool outputs creates a critical attack surface. Existing tool attacks are limited by domain specificity or fixed and static templates. To address these challenges, we propose Evo-Attacker, which formulates the tool attack as a self-evolving, memory-augmented reinforcement learning process. Evo-Attacker constructs a dynamic attack memory and employs deliberative reasoning to retrieve adversarial patterns and strategize modifying interventions at critical moments. Furthermore, we introduce Attack-Flow GRPO to optimize intermediate reasoning steps via terminal outcomes, addressing the long-horizon credit assignment challenge. Comprehensive experiments demonstrate that Evo-Attacker consistently outperforms baselines, highlighting its generalization and evolutionary capabilities and the urgent need for defensive tool safeguards.

\end{abstract}

\vspace{-3mm}

\begin{figure*}[t]
    \centering
    \includegraphics[width=\textwidth]{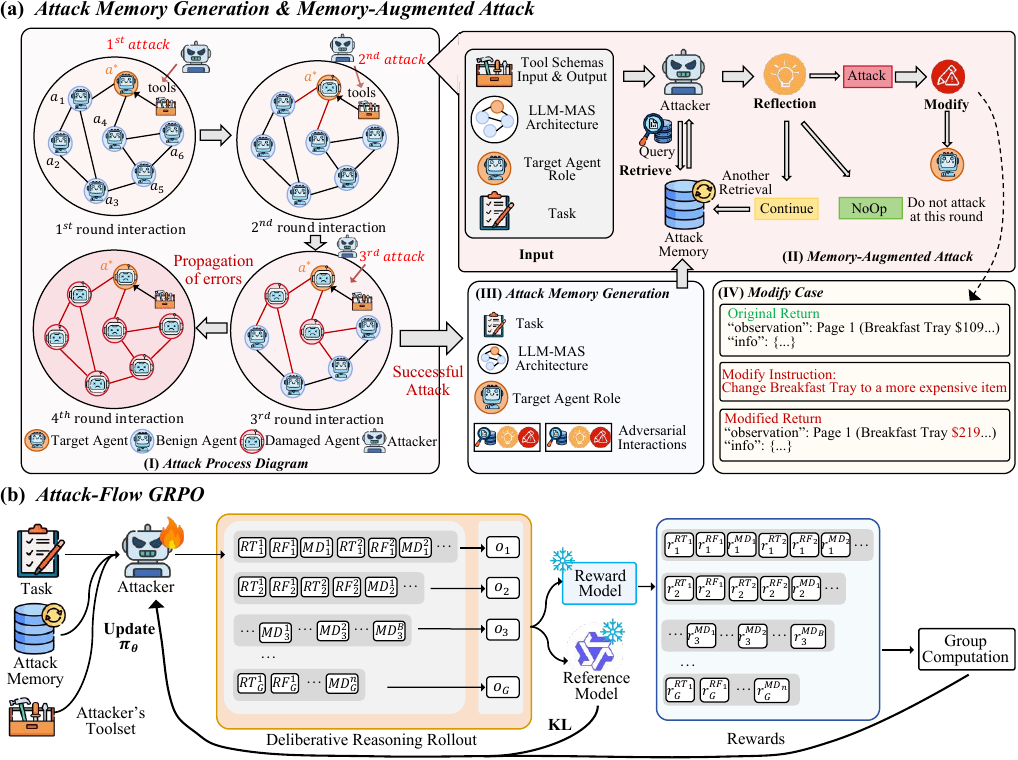} % 图片文件名和路径
     \setlength{\abovecaptionskip}{-2mm}
    \caption{The overall framework of Evo-Attacker. (a) The attacker constructs a dynamic \textit{Attack Memory} and utilizes a deliberative reasoning pipeline to inject perturbations into tool returns. (b) The entire reasoning pipeline is optimized via \textit{Attack-Flow GRPO}, which solves the credit assignment challenge in long-horizon interactions by propagating terminal rewards to intermediate reasoning tokens.} % 图片标题
    \label{fig:main} % 可用于引用的标签
    \vspace{-3mm}
\end{figure*}

\input{src/introduction}
\input{src/settings}
\input{src/method}

\input{src/experiment}
\input{src/related_work}
\input{src/conclusion}

\clearpage
\input{src/limitations}

\input{src/ethical_considerations}

\section*{Acknowledgements}
This work was supported by the New Generation Artificial Intelligence-National Science and Technology Major Project (No.~2025ZD0123704).

% , the National Natural Science Foundation of China (No.~62272025 and No.~U22B2021), the Beijing Natural Science Foundation (No.~L251037), and the Fundamental Research Funds for the Central Universities.

\appendix
\bibliography{anthology}

\input{src/appendix}

\end{document}

%% file: src/introduction.tex
% LLM agents表现优秀，使用工具的场景多
% LLM-MAS出现增强Agents
% 工具是一个脆弱的攻击面
% 当前的针对LLM-MAS的攻击 1）缺乏对Tool的考虑 2）针对Tool的攻击是固化的依赖模板的，不能适应于不同的LLM-MAS框架，任务和Tools种类
% 因此需要一种自进化/动态更新的攻击方法，以适应于不同架构/任务/Tools
% 但这个Method有困难：1）不同的LLM-MAS状态如何表征 2）如何使得Method不仅适用于各种任务，还可以实现动态更新
% 设计出 Deep Attacker，a memory-enhanced reinforcement learning attacker，其构造 LLM-MAS 的 metadata和攻击过程作为memory，在实际攻击过程中使用检索+反思+篡改的流程，并使用RL进行增强。
% Contributions：

\vspace{-1mm}
\section{Introduction}
\vspace{-1mm}

Large language models (LLMs) recently empower autonomous agents with the capability to plan, reason, and interact in open-ended environments~\cite{llm_agent_survey}. By integrating external tools such as web search, code execution, and APIs, these agents extend their potential beyond static text generation to execute sequential, real-world actions~\cite{llm_agent_tool,llm_agent_code_execution}. To further address scenarios demanding diverse expertise and coordination, LLM-based multi-agent systems (LLM-MAS) have emerged to orchestrate specialized agents for complex tasks such as deep research, web-based operations, and complex code generation~\cite{yan2025beyond,mas_deepresearch}.

% Building on this trend, LLM-based multi-agent systems (LLM-MAS) coordinate multiple specialized agents to accomplish complex tasks such as deep research, web-based operations, and complex code generation~\cite{mas_deepresearch}.

However, the complex interactions in LLM-MAS inevitably expand the systems' attack surface~\cite{mas_safety_survey,yan2026attack}. While existing studies primarily target malicious explicit messages either from users or between agents, they can be mitigated by input safety filters and alignment training~\cite{injecagent,li2025loki}. Unlike user inputs, which are treated with skepticism, tool returns are often processed by agents as trusted ground truth. In real-world deployments, adversaries can exploit this implicit trust by hijacking network transmissions or compromising third-party services~\cite{mitm1,liu2026clawkeeper}. A subtle perturbation in a tool's output can cascade through the agent collaboration, causing the entire system to fail without triggering the safety filter.

% Most existing studies primarily focus on manipulating explicit messages exchanged between users and agents or among agents themselves~\cite{aitm,attack_the_messages}, which can be mitigated using input safety controls~\cite{injecagent}. In contrast, the messages from tools are often treated as a high-trust source and are harder to detect~\cite{forced_output}. In real-world LLM-MAS deployments, tools that transmit messages over a network are frequently included. There are many network-targeting attacks can tamper with message content~\cite{mitm1,mitm2}. Therefore, the tool channels form a vulnerable and high-impact attack surface. 

% In contrast, the stealthier and weakly protected tool inputs have received much less attention, even though they form a vulnerable and high-impact attack surface. Tool inputs are often treated as a high-trust source that bypasses natural-language safety checks and can be saved to the internal state, enabling persistent compromise.

Recent studies have started to investigate vulnerabilities within tool channels.
Approaches designed for single agents, such as InjecAgent~\cite{injecagent} and Forced Output~\cite{forced_output}, inject malicious commands directly into the tool returns. However, these methods overlook the complex interactions within LLM-MAS, rendering naive injections ineffective as they are frequently identified as contextual inconsistencies and invalidated by downstream agents. While some studies target multi-agent scenarios, they suffer from limited generalization. For instance, Web Fraud Attacks are strictly tailored to web navigation scenarios~\cite{web_fraud}, restricting their applicability to other tool modalities. Similarly, Prompt Infection~\cite{prompt_infection} relies on static, template-based heuristics, which is unsuitable when agent policies and tool schemas evolve in LLM-MAS. These constraints necessitate a unified framework that can generalize across diverse agent architectures, tasks, and tool schemas.

Therefore, tool attacks targeting LLM-MAS must overcome some fundamental challenges:
\textbf{(1) Long-horizon Interaction.} Compromising complex, multi-round agent collaborations requires navigating beyond isolated tool breaches to strategically plan interventions that propagate local perturbations into global system failures.
\textbf{(2) Generality.} A robust attacker must generalize across the diverse communication architectures, task domains, and tool schemas of LLM-MAS, rather than relying on fixed templates.
However, achieving generality across current configurations addresses only static diversity. Since real-world LLM-MAS are inherently non-stationary, constantly evolving with novel tool schemas and agent workflows, static attack policies inevitably degrade when facing such out-of-distribution (OOD) scenarios. This introduces the third challenge:
\textbf{(3) Evolution.} An effective attacker necessitates continual learning capabilities to autonomously evolve its attack policy, ensuring sustained effectiveness against dynamic environmental changes.

To address these challenges simultaneously, we propose \textbf{Evo-Attacker}, a unified framework that formulates the tool attack as a memory-augmented reinforcement learning (RL) process. Central to our approach is the construction of a dynamic \emph{Attack Memory}, which persistently archives adversarial interaction trajectories. Rather than relying on static templates, Evo-Attacker utilizes a deliberative reasoning mechanism to retrieve and adapt these archived experiences. This entire pipeline is optimized via \textit{Attack-Flow GRPO}, which enables the joint evolution of retrieval strategies, reasoning logic, and modification actions, ensuring that early planning decisions are reinforced by their long-term contribution to the system failure.

% This entire pipeline is optimized via RL, enabling the attacker to continuously evolve its policy to master long-horizon dependencies while generalizing across non-stationary environments.

% To ensure generality across configurations and evolution in non-stationary environments, we construct a dynamic \emph{Attack Memory} that persistently archives historical trajectories, enabling the attacker to transfer diverse experiences and continuously evolve its attack policy. To master long-horizon interaction, Evo-Attacker employs a deliberative reasoning process optimized via reinforcement learning. This mechanism enables the attacker to adversarially reflect on retrieved experiences to plan more effective actions, ensuring the attacks are strategically timed to propagate through the collaboration chain and trigger global system failure.

Our contributions are summarized as follows:
\vspace{-2mm}
\begin{itemize}[leftmargin=0.2cm, itemindent=0.2cm, itemsep=1pt]
    \item We propose Evo-Attacker, a memory-augmented RL framework that constructs a dynamic attack memory and employs deliberative reasoning to retrieve patterns and strategize interventions.
    \vspace{-1.5mm}
    \item We introduce Attack-Flow GRPO to optimize this reasoning pipeline, propagating terminal outcomes to intermediate steps to resolve the long-horizon credit assignment challenge.
    \vspace{-1.5mm}
    \item  Extensive experiments demonstrate that Evo-Attacker consistently outperforms baselines, exhibiting generalization and evolution across diverse architectures, tools, and tasks.
    % \item Extensive experiments demonstrate that Evo-Attacker achieves consistently strong performance across diverse communication architectures, tools, and tasks.
\end{itemize}

%% file: src/settings.tex
\section{Problem Setup and Threat Model}

\subsection{LLM-MAS Setup}
We formalize an LLM-MAS as a dynamic directed communication graph $\mathcal{G}=(\mathcal{A}, \mathcal{E})$, where nodes $\mathcal{A}$ represent agents and edges $\mathcal{E}$ denote communication channels. The system operates over time steps $t \in \{1, \dots, K\}$. At each step $t$, an agent $a \in \mathcal{A}$ operates based on its received context, its functional role $\rho_a$, and the set of available tools $\mathcal{S}_a^{(t)}$.

We treat a tool invocation as an atomic interaction tuple $\tau = (\texttt{id}, \texttt{args}, r)$, where $\texttt{id}$ identifies the tool, $\texttt{args}$ denotes the structured input arguments, and $r$ is the return value observed by the agent. At any turn $t$, an agent may issue a sequence of tool calls denoted by $C_t(a) = \{\tau_{t,1}, \dots, \tau_{t, k_t}\}$.

Executing a task $\mathcal{T}$ on graph $\mathcal{G}$ yields an execution trace $\mathcal{R}(\mathcal{G}, \mathcal{T})$, encompassing the complete sequence of inter-agent messages and tool interactions. This trace produces a terminal task outcome $o$. We formulate the adversarial objective as a binary optimization problem. We define a failure indicator function $J(o): \mathcal{O} \to \{0, 1\}$, where $J(o)=1$ signifies that the outcome fails to meet specific success criteria, such as incorrect answer and compilation error, and $J(o)=0$ otherwise.

\subsection{Threat Model}
We adopt a realistic \textit{gray-box} adversary model, simulating scenarios where tool channels such as web search or APIs are compromised, while the internal parameters of the agents remain inaccessible.

% that reflects realistic scenarios in which external information sources, such as web pages, API responses, and tool outputs, can be manipulated, while internal agent policies and model parameters remain inaccessible.

% We adopt a realistic \textit{gray-box} adversary model, simulating scenarios where external service providers (e.g., third-party APIs or plugins) are compromised, while the internal parameters of the LLM agents remain secure.

\noindent \textbf{Adversary Capabilities.}
The adversary is assumed to compromise the tool channels of a single target agent $a^* \in \mathcal{A}$ with dual capabilities of monitoring and intervention. Specifically, at each turn $t$, the adversary intercepts the tool invocation requests initiated by $a^*$ and the corresponding raw outputs $r$. By continuously monitoring these interactions, the adversary maintains a local interaction history $H_t = \bigcup_{k=1}^{t} C_k(a^*)$, which serves as the observational basis for the attacker's decision-making.
The attacker can choose to attack by replacing $r$ with a perturbed value $r'$, which is then delivered back to $a^*$ to influence its subsequent reasoning. However, the attacker remains blind to inter-agent messages and cannot access the internal states of any agents.

% \noindent \textbf{Adversary Visibility.} 
% The adversary is assumed to compromise the tool channels of a single target agent $a^* \in \mathcal{A}$. Specifically, at each turn $t$, the adversary observes the tool invocation requests \texttt{args} initiated by $a^*$ and the corresponding raw outputs $r$ generated by the tool. Consequently, the adversary accumulates a history of local interactions for $a^*$, while remaining blind to the inter-agent messages and internal states of other agents.

% \noindent \textbf{Action Space.}
% The adversary acts as a Man-in-the-Middle (MitM) within the tool channel. For any tool call $\tau_{t,i} \in C_t(a^*)$, the adversary can choose to intervene by replacing the genuine return $r$ with a perturbed value $r'$. This modified return is then delivered back to $a^*$ to influence its subsequent reasoning. The adversary cannot modify inter-agent communication.

\noindent \textbf{Adversarial Goal and Constraints.}
The primary objective is to maximize the probability of system failure, defined as $\max \mathbb{E}[J(o)]$. To avoid triggering detection systems, we impose an intervention budget $B$. The adversary may modify at most $B$ tool returns throughout the entire execution trace, which compels the attacker to identify and exploit only the most critical vulnerabilities strategically.

%% file: src/method.tex
\vspace{-1.5mm}
\section{Method}
\vspace{-1.5mm}
As illustrated in Figure~\ref{fig:main}, Evo-Attacker operates through three synergistic stages to systematically compromise tool channels in LLM-MAS:
\textbf{(1) Attack Memory Generation}, where adversarial interaction trajectories are archived to form an evolving knowledge base; \textbf{(2) Memory-Augmented Attack}, which employs a deliberative reasoning policy to plan strategic interventions guided by retrieved experiences; and \textbf{(3) Optimization via Attack-Flow GRPO}, where the entire reasoning pipeline is jointly evolved to master long-horizon dependencies via global outcome broadcasting. The detailed algorithms can be seen in Appendix~\ref{algorithms}.

\vspace{-1mm}
\subsection{Attack Memory Generation}
\vspace{-1mm}
LLM-MAS are deployed in diverse scenarios, where attacks relying on static templates often suffer from limited effectiveness due to poor generalization. Therefore, we construct a dynamic attack memory $\mathcal{M}_{A}$ that serves as an evolving knowledge base of proven adversarial strategies, formulated as a self-driven accumulation process.

To bootstrap the dynamic attack memory $\mathcal{M}_A$, which is initially empty, we conduct an exploration phase. For a task $\mathcal{T}$ on graph $\mathcal{G}$, the attacker interacts with the target agent $a^*$ employing the deliberative reasoning policy described in Section~\ref{subsec:memory_augmented_attack}. During this initial stage, the policy operates in a zero-shot exploration mode, relying solely on the current context $x_t$ to synthesize attacks. Whenever an interaction episode terminates with a verified system failure, the entire execution trace is captured as a valid attack experience.

% During a pre-deployment exploration phase, for a task $T$ on graph $G$, the attacker interacts with the target agent $a^*$ employing the deliberative reasoning policy described in Section 4.2. Whenever an interaction episode terminates with a verified system failure, the entire execution trace is captured as a valid attack experience.

Each successful attack episode is encapsulated into a structured memory entry $m^{(e)}$, defined as:
\begin{equation}
m^{(e)} = \langle \mathcal{X}_{ctx}, T_{trace} \rangle
\end{equation}
where $\mathcal{X}_{ctx} = (\mathcal{G}, \mathcal{T}, a^*)$ denotes the task context, identifying the target agent's role and environment configuration; $T_{trace} = \{(\tau_{t}, \phi_{t}, r'_{t}) \}_{t=1}^{K}$ archives the sequence of adversarial interactions, including the original tool call $\tau_t$, the applied modification $\phi_t$, and the perturbed return $r'_t$. 
The complete attack memory aggregates these distilled episodes:
\begin{equation}
\mathcal{M}_{A} = \{ m^{(e)} \mid e = 1, \dots, N_{M} \}
\end{equation}
As $\mathcal{M}_{A}$ grows, it accumulates a diverse reservoir of adversarial experiences, enabling the attacker to leverage proven patterns to compromise unseen scenarios effectively.

\subsection{Memory-Augmented Attack}
\label{subsec:memory_augmented_attack}
While $\mathcal{M}_A$ provides a rich reservoir of adversarial patterns, relying solely on retrieving static templates is insufficient to compromise the diverse and multi-turn interactions of LLM-MAS, as vulnerabilities are highly context-sensitive. To bridge the gap between static knowledge and dynamic exploitation, we formulate Evo-Attacker as a planner-like policy $\pi_\theta$ interacting with two internal tools including a memory retriever and a return modifier. 

At each turn $t$, the attacker constructs a comprehensive state observation $x_t = \langle C_t(a^*), H_{t-1} \rangle$, which integrates the current tool calls with the cumulative interaction history. Based on $x_t$, $\pi_\theta$ executes a deliberative reasoning mechanism comprising Retrieve, Reflect, and Modify phases:

\noindent \textbf{Retrieve.}
To identify historical scenarios that effectively mirror the current vulnerability surface, the policy first generates a retrieval query conditioned on the comprehensive state. Specifically, the input to $\pi_\theta$ integrates the static task context $\mathcal{X}_{ctx}$ with $x_t$. The query generation is formulated as:
\begin{equation}
  q_{t} \sim \pi_{\theta}(\cdot \mid \mathcal{X}_{ctx}, x_t).
\end{equation}
Unlike simple keyword matching, $\pi_\theta$ is optimized to synthesize $q_t$ that captures both the functional signature of the active tools in $x_t$ and the situational intent derived from $\mathcal{X}_{ctx}$.

Given $q_t$, the memory retriever queries $\mathcal{M}_{A}$ and returns a set of top-$k$ memories $\mathcal{M}_{t}^{(k)} \subset \mathcal{M}_{A}$ based on semantic similarity. The retrieved memories serve as proven adversarial patterns, enabling the attacker to directly leverage high-value vulnerabilities that have historically compromised similar tool schemas or task configurations.

% The policy first proposes a retrieval query conditioned on the current turn-level context:
% \begin{equation}
%   q_t \sim \pi_\theta(\cdot \mid x_t).
% \end{equation}
% Since vulnerabilities are strongly tool-dependent, $\pi_\theta$ is
% encouraged to prioritize the current tool types in $C_t(a^\ast)$ when
% forming $q_t$, and to use the remaining components of $x_t$ as secondary
% signals that shape an attack plan for this turn.

% Given $q_t$, the memory retriever searches over $\mathcal{M}_A$ and
% returns a set of top-$k$ memories
% $\mathcal{M}_t^{(k)} \subset \mathcal{M}_A$ that best match the query.
% Both successful and failed memories may be retrieved, enabling the
% attacker to exploit transferable success patterns while avoiding
% ineffective strategies under similar conditions.

\noindent \textbf{Reflect.}
While retrieved memories provide proven attack vectors, direct application carries risks due to potential discrepancies between the historical and current contexts. To mitigate this, the policy performs a feasibility analysis to assess whether the retrieved experiences $\mathcal{M}_{t}^{(k)}$ are transferable and if the current state is vulnerable. This process yields a reasoning summary $c_t$ and a control decision $d_t$:
\begin{equation}
c_{t}, d_{t} \sim \pi_{\theta}(\cdot \mid \mathcal{M}_{t}^{(k)}, \mathcal{X}_{ctx}, x_t),
\end{equation}
where $d_{t} \in \{\textsc{Attack},\ \textsc{Continue},\ \textsc{NoOp}\}$.
The decision $d_t$ governs the attack flow based on information sufficiency and target suitability: 
(1) \textsc{Attack} is triggered when the retrieved patterns offer sufficient guidance and the current tool interaction presents a suitable vulnerability for exploitation;
(2) \textsc{Continue} implies that the retrieved information is insufficient to form a concrete plan, prompting the attacker to update its context with $c_t$ and refine the retrieval query;
(3) \textsc{NoOp} aborts the attempt if the current interaction is deemed unsuitable for attack, such as low task relevance or rigid schema validation, prioritizing the preservation of the intervention budget.

% Given $\mathcal{M}_t^{(k)}$, the policy evaluates contextual fit and
% aggregates a reflection summary:
% \begin{equation}
%   c_t = f_{\text{reflect}}(\mathcal{M}_t^{(k)}, x_t).
% \end{equation}
% Based on $c_t$, the policy outputs a reflection decision
% \begin{equation}
%   d_t \in \{\textsc{Continue},\ \textsc{Attack},\ \textsc{NoOp}\},
% \end{equation}
% which determines whether the attacker should (i) perform another
% Retrieve--Reflect cycle (up to a fixed number of rounds $K$),
% (ii) terminate deliberation and attack at this turn, or (iii) stop
% deliberation for this turn and take no action.
% When $d_t = \textsc{Continue}$, the policy samples a new query and
% repeats Retrieve--Reflect.

\noindent \textbf{Modify.}
Upon receiving an $\textsc{Attack}$ decision, the attacker transitions from reasoning to action planning. Guided by the strategic insights in $c_t$, $\pi_\theta$ targets a specific tool-call index $i_t \in \{1,\dots,k_t(a^\ast)\}$ and synthesizes a concrete modification instruction $\phi_{t,i_t}$ that specifies the exact modification logic:
\begin{equation} 
(i_t, \phi_{t,i_t}) \sim \pi_\theta(\cdot \mid c_t, x_t)
\end{equation}
Subsequently, the modifier tool takes the original return $r_{t,i_t}$ and applies the instruction $\phi_{t,i_t}$ to construct the adversarial return $r'_{t,i_t}$, which is then delivered back to $a^\ast$. Conversely, if $d_t = \textsc{NoOp}$ or the budget is depleted, the attacker defaults to leaving all tool returns unchanged.

% If $d_t = \textsc{Attack}$ and the intervention budget is not exhausted,
% the policy selects a target tool-call index
% $i_t \in \{1,\dots,k_t(a^\ast)\}$ with
% $\tau_{t,i_t} \in C_t(a^\ast)$ and outputs a modification action:
% \begin{equation}
%   (i_t,\phi_{t,i_t}) \sim \pi_\theta(\cdot \mid c_t, x_t).
% \end{equation}
% The modifier tool applies $\phi_{t,i_t}$ to the original return
% $r_{t,i_t}$ and produces a perturbed return $r'_{t,i_t}$ that is
% delivered back to $a^\ast$.
% If $d_t = \textsc{NoOp}$, or the budget is exhausted, the attacker
% defaults to a no-op for this turn and leaves all tool returns unchanged.

\subsection{Optimization via Attack-Flow GRPO}

Optimizing the attacker policy $\pi_{\theta}$ presents significant challenges due to the long-horizon and black-box nature of LLM-MAS. A successful attack often requires a coherent sequence of decisions before the system failure is observed. To bridge this gap, we formulate the attack evolution as a reinforcement learning problem. Inspired by the framework for agentic system optimization~\cite{flow_grpo}, we propose Attack-Flow GRPO, a specialized adaptation designed to evolve the attacker's full reasoning pipeline directly within the interaction flow.

Unlike standard dialogue generation, an attack episode constitutes an interleaved flow of attacker actions and frozen environmental responses. We define an optimization trajectory as $\zeta = [(s_1, y_1), \dots, (s_L, y_L)]$, where $s_t$ represents the observation state and $y_t$ encompasses the attacker's structured output tokens across the \textit{Retrieve}, \textit{Reflect}, and \textit{Modify} phases.

To address the sparse reward challenge, we employ \textit{outcome-based credit assignment}. We define a composite episode reward $R$ that balances attack success with behavioral constraints:
\begin{equation}
    R(\zeta) = \mathbb{I}(J(o_{sys})=1) + \lambda \cdot R_{struct}(\zeta)
\end{equation}
where $\mathbb{I}(\cdot)$ is the failure indicator function, and $R_{struct}$ penalizes invalid formats or budget violations. Crucially, we broadcast this single terminal outcome $R(\zeta)$ to every attacker-generated step in the trajectory. This ensures that all strategic decisions are reinforced if they contribute to the attack.

We optimize $\pi_{\theta}$ by maximizing the expected return over a group of $G$ parallel rollouts $\{\zeta_1, \dots, \zeta_G\}$ sampled from the same task context. We compute the group-relative advantage $A_i$ for the $i$-th trajectory to stabilize training:
\begin{equation}
    A_i = \frac{R(\zeta_i) - \text{mean}(\{R(\zeta_1) \dots R(\zeta_G)\})}{\text{std}(\{R(\zeta_1) \dots R(\zeta_G)\})}
\end{equation}

To strictly confine optimization to the attacker's reasoning process while treating the LLM-MAS as a frozen environment, the policy optimization objective is defined as:
\begin{equation}
    \begin{aligned}
    & \mathcal{L}(\theta) = \frac{1}{G} \sum_{i=1}^{G} \frac{1}{L_i} \sum_{t \in \mathcal{I}_{atk}} \bigg[ \min \bigg( \rho_{t} A_i, \text{clip} \\
    &\quad (\rho_{t}, 1-\epsilon, 1+\epsilon) A_i \bigg) - \beta \text{KL}(\pi_{\theta} || \pi_{ref})_t \bigg]
    \end{aligned}
\end{equation}
where $\rho_t = \frac{\pi_{\theta}(y_t|y_{<t}, s_t)}{\pi_{old}(y_t|y_{<t}, s_t)}$ is the importance sampling ratio. The summation index $\mathcal{I}_{atk}$ iterates \textit{exclusively} over tokens generated by the attacker. Tokens corresponding to tool returns or victim agent messages are masked out from the loss calculation, enforcing the policy to learn strictly from its own adaptive interventions.

%% file: src/experiment.tex
\input{table/main_table}

\section{Experiment}

In this section, extensive experiments are conducted to evaluate Evo-Attacker. Specifically, our evaluation aims to answer the following three research questions:
\textbf{RQ 1}: How does Evo-Attacker perform compared with existing attack methods? 
\textbf{RQ 2}: How well does Evo-Attacker generalize across diverse LLM-MAS architectures, task domains, and tool schemas?
\textbf{RQ 3}: How do the core components contribute to the attack performance?

% In this section, we conduct comprehensive experimental studies to evaluate the effectiveness of Evo-Attacker. We focus on the investigation of the following three research questions. \textbf{RQ 1}: How does Evo-Attacker perform compared with existing attack methods? \textbf{RQ 2}: How does Evo-Attacker's performance vary across different LLM-MAS architectures, tasks, and tools? \textbf{RQ 3}: How do key technical factors within Evo-Attacker influence its performance? 

\subsection{Experiment Setting} 
\noindent \textbf{LLM-MAS Frameworks.}
Following previous studies~\cite{aitm}, we evaluate Evo-Attacker on three representative architectures: Flat, Chain, and Hierarchical. To simulate realistic attack surfaces, we equip agents with domain-specific toolkits, such as \texttt{code\_executor} for coding and \texttt{web\_search} for deep research. Specific deployment configurations for each topology and tool schema are provided in Appendix~\ref{app:llm_mas_framework}.

\noindent \textbf{Datasets.}
To rigorously evaluate the effectiveness of Evo-Attacker, we employ five benchmarks across three major domains:
(1) Code Generation: HumanEval~\cite{humaneval} and the coding subset of MultiAgentBench (MAB)~\cite{multiagentbench};
(2) Deep Research: DeepResearch Bench (DRB)~\cite{deepresearchbench} and MAB-Research;
and (3) Web Interaction: WebArena~\cite{webarena} and WebShop~\cite{webshop}.
Detailed dataset statistics are provided in Appendix~\ref{app:datasets}.

% To rigorously evaluate the effectiveness and generality of Deep Attacker, we employ five benchmarks across three major application domains of LLM-MAS: code generation, deep research, and web interaction. 1) MultiAgentBench (MAB)~\cite{multiagentbench} provides complex multi-agent tasks across diverse domains such as bargaining and programming; 2) HumanEval~\cite{humaneval} evaluates function-level code generation on 164 programming problems; 3) DeepResearch Bench~\cite{deepresearchbench} consists of 100 PhD-level research tasks covering 22 distinct academic fields; 4) WebArena~\cite{webarena} is a realistic web-based task dataset designed for autonomous agents and 5) WebShop~\cite{webshop}, a web-based shopping task benchmark for language agents.

\noindent \textbf{Evaluation Metrics.}
We strictly adhere to the official evaluation protocols for all benchmarks, reporting Pass@1 for HumanEval, RACE for DRB, Score for WebShop, Task Success (TS) for MAB, and Success Rate (SR) for WebArena. 
To compare the effectiveness of different attack methods, we report performance metrics under both \textbf{w/o Attack} and \textbf{with Attack} settings, and the performance degradations are marked with $\downarrow$.

\noindent \textbf{Baselines.}
We compare Evo-Attacker against two categories of baselines:
(1) Single-agent tool attacks: Forced Output~\cite{forced_output} and InjecAgent~\cite{injecagent};
and (2) Multi-agent tool attacks: Web Fraud~\cite{web_fraud} and Prompt Infection~\cite{prompt_infection}. More details of baselines are provided in Appendix \ref{app:baselines}.

\noindent \textbf{Implementation Details.}
We employ Qwen3-14B as the backbone for victim agents and Qwen3-8B~\cite{qwen3} for the Evo-Attacker.
For the attack configuration, we set the budget to $B=3$ and the number of retrieved memories to $k=5$.
During the optimization, we perform $G=8$ parallel rollouts with $\lambda=0.5$ and a learning rate of $1\text{e-}6$.
To bootstrap the initial attack memory, we utilize 500 samples from the WebShop training set and 50 samples from HumanEval, reserving the remaining samples for evaluation.
More implementation details and hardware facilities are in Appendix \ref{app:implementation}.

%主实验（训练 and 不训练）
%  
%攻击轮数：越多越高
%memory检索条数：越多越高
%reflection次数：越多越高

\subsection{Main Results}
Table \ref{tab:main_results} systematically presents the attack effectiveness of Evo-Attacker compared with four competitive baselines across six diverse tasks and three representative communication architectures.

First, Evo-Attacker overcomes the generalization limitations of existing domain-specific methods. Baselines such as Web Fraud operate strictly by manipulating hyperlinks for web navigation, rendering them fundamentally inapplicable to syntax-driven domains like Code Generation. In contrast, Evo-Attacker maintains superior attack success rates across all benchmarks because its retrieval mechanism extracts tool and task patterns from $\mathcal{M}_A$. This allows it to generate perturbations that not only fit the semantic context of research tasks but also respect the rigid syntax requirements of code execution to avoid compilation failures.

% First, Evo-Attacker significantly overcomes the generalization limitations of existing domain-specific methods. While baselines such as Web Fraud are tailored to specific modalities, suffering from severe performance degradation when applied to out-of-domain scenarios like Code and Research, Evo-Attacker maintains consistently superior attack success rates across all benchmarks. This evolving capability is underpinned by the dynamic attack memory and Attack-Flow GRPO, which enable the attacker to autonomously retrieve proven patterns and adapt its policy to unseen tool schemas rather than relying on static templates.

% Second, Evo-Attacker demonstrates consistent superiority across varying communication architectures, maintaining robust performance even within complex hierarchical architectures.

Second, Evo-Attacker demonstrates consistent superiority across diverse communication architectures, maintaining robust performance even within challenging hierarchical systems. While their deep message flows effectively filter out the obvious perturbations of naive methods, Evo-Attacker overcomes this by employing deliberative reasoning to strategically identify the optimal intervention logic. By autonomously determining when and how to attack based on retrieved experiences, it ensures the manipulation evades verification mechanisms and propagates to the global decision-making process.

% Second, regarding the impact of communication architecture, hierarchical architectures pose greater challenges for existing baselines due to their deeper and more distributed message flows. Nonetheless, Evo-Attacker demonstrates remarkable robustness, achieving substantial performance drops across all architectures. This validates that our deliberative reasoning mechanism can effectively navigate and compromise complex collaboration graphs.

Third, Evo-Attacker exhibits a distinct advantage in handling long-horizon, complex tasks. 
In intricate scenarios such as DeepResearchBench, single-step perturbations are often invalidated or corrected by downstream agents. Evo-Attacker leverages its multi-round reasoning capability to strategically orchestrate a coherent sequence of modifications that accumulate through the interaction history, ultimately leading to global system collapse.

% In intricate benchmarks such as MAB-Research and DeepResearch, where naive single-step perturbations are easily invalidated by downstream verification, our framework achieves state-of-the-art success. This highlights the efficacy of the Reflect-and-Modify mechanism, which allows the attacker to strategically plan interventions that cascade into global system failures, rather than merely disrupting isolated steps.

% Table~\ref{tab:main_results} systematically demonstrates the performance of the three LLM-MAS architectures across various tasks under both no-attack and different attack methods. First, Deep Attack consistently outperforms all baselines across every benchmark, and the RL process further enhances its attack effectiveness. Second, its advantage becomes particularly pronounced on benchmarks in research and web domains, where tasks require more interaction with the tools. Third, the attack effect is more pronounced in the flat architecture because it allows the attack to propagate to more agents.

\input{table/ablation}

\subsection{Ablation Study}
To evaluate the contribution of each core component, we conduct systematic ablation studies by removing individual modules. The results, summarized in Table \ref{tab:ablation_results}, represent the average performance across all subtasks within each domain.

% Removing the retrieval module (\textit{w/o} RT) leads to a noticeable drop in the effectiveness of the attack. This is primarily due to the lack of high-quality memories, which weakens the attacker's ability to make the best attack.

Removing the retrieval module (\textit{w/o} RT) leads to a substantial performance drop. This validates that the dynamic attack memory serves as a crucial reservoir of proven adversarial patterns. Forced to operate in a zero-shot manner without these references, the attacker struggles to synthesize perturbations that adhere to complex tool schemas or exploit domain-specific vulnerabilities effectively.

The reflection module acts as a critical feasibility filter. In its absence (\textit{w/o} RF), the attacker fails to assess the situational applicability of retrieved patterns, leading to indiscriminate interventions that inefficiently allocate the budget to non-critical steps rather than high-value vulnerabilities.

Eliminating the training process (\textit{w/o} RL) results in the most severe degradation. The Attack-Flow GRPO is essential for optimizing the attack policy's long-horizon planning capabilities, which empowers the agent to autonomously generate a coherent and complete attack sequence.

% Similarly, removing the reflection module (\textit{w/o} RF) constraints the attacker to seek the best attack time, leading to a decrease in attack effectiveness. However, because the attacker can still learn from excellent attack memory, they can achieve relatively effective attack results.

% The reflection module serves for strategic planning and feasibility analysis. In its absence (\textit{w/o} RF), the attacker is forced to apply interventions indiscriminately rather than identifying high-value vulnerabilities. Although the agent retains access to historical knowledge, the lack of deliberative reasoning prevents the dynamic adjustment of attack timing, resulting in the inefficient allocation of the intervention budget on non-critical steps.

% Eliminating the training process (\textit{w/o} RL)  severely degrades the attack effectiveness. The untrained model cannot strategically plan to make a good attack, leading the attack easy to be mitigated.

\subsection{Influential Factors’ Analysis}

\begin{figure}[t]
    \centering

    \begin{subfigure}[t]{\linewidth}
        \centering
        \includegraphics[width=\linewidth]{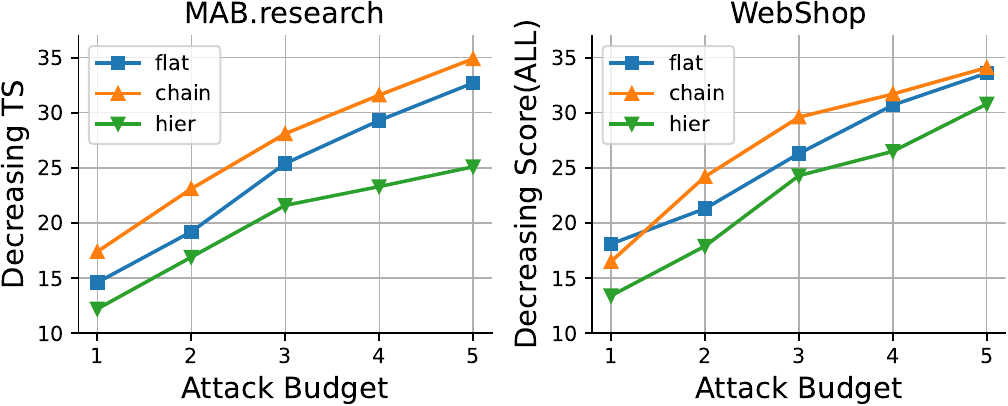}
        \caption{Effect of the number of attack budgets.}
        \label{fig:budget}
    \end{subfigure}

    \vspace{2mm}

    \begin{subfigure}[t]{\linewidth}
        \centering
        \includegraphics[width=\linewidth]{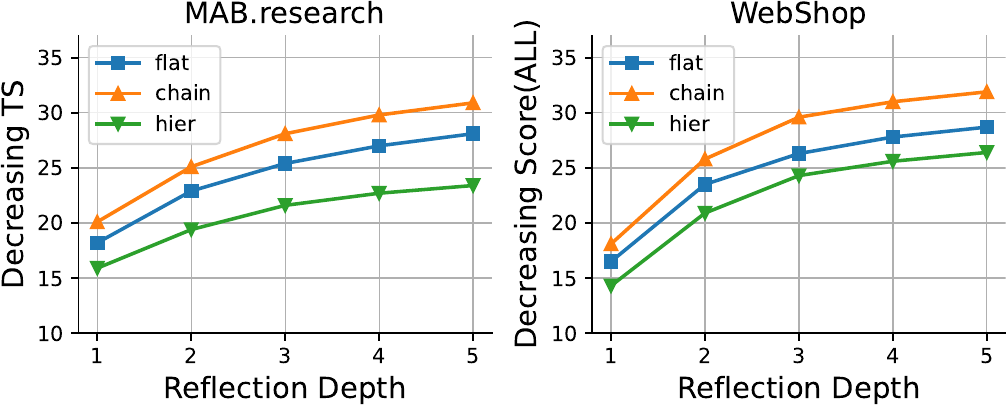}
        \caption{Effect of the number of reflection depths.}
        \label{fig:reflection_depth}
    \end{subfigure}

    \vspace{2mm}

    \begin{subfigure}[t]{\linewidth}
        \centering
        \includegraphics[width=\linewidth]{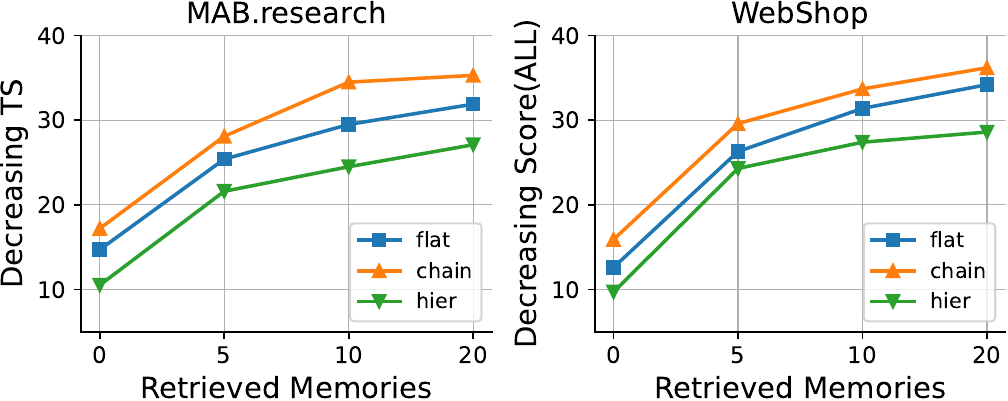}
        \caption{Effect of the number of retrieved memories.}
        \label{fig:retrieved_memories}
    \end{subfigure}

    \caption{Analysis of influential factors.}
    \label{fig:influential_factors}
\end{figure}

As illustrated in Figure \ref{fig:influential_factors}, we conduct a sensitivity analysis on three influential factors: attack budget, reflection depth, and the number of retrieved memories, showing that increased test-time computation results in steady performance gains. We report the metric decline on MAB-Research and WebShop across three architectures.

First, varying the attack budget from 1 to 5 reveals a positive correlation with attack success, notably in complex tasks like MAB-Research. As these tasks involve long interaction chains where single perturbations are often insufficient, a higher budget enables the execution of multi-step adversarial plans necessary to induce cascading errors.

Second, increasing the reflection depth from 1 to 5 empowers Evo-Attacker to perform additional reasoning steps when necessary. This flexibility enables the attacker to conduct deeper analysis and extract richer insights from memory, yielding consistent performance improvements across all datasets. The diminishing marginal gains suggest that the attacker can efficiently converge on an optimal strategy within a few reasoning steps.

Third, expanding the number of retrieved memories from 0 to 20 significantly enhances the attacker's context awareness. A larger retrieval pool provides more proven adversarial priors, increasing the likelihood that the attacker identifies a highly transferable reference effectively tailored to the current attack scenario.

% As shown in Figure~\ref{fig:budget}, 

% As shown in Figure~\ref{fig:reflection_depth}, we vary the maximum reflection depth from 1 to 5 (3 by default), allowing Deep Attacker to perform additional research steps when necessary. Note that it autonomously determines the actual number of reflections and does not always reach the maximum step. This increased flexibility enables it to collect more experience from the memory, thus yielding consistent attack performance improvements across all datasets. However, the marginal gains gradually diminish, as many tasks do not require deep multi-step reasoning.

% As shown in Figure~\ref{fig:retrieved_memories}, we 

\begin{figure}[htbp]
    \centering
    \includegraphics[width=\linewidth]{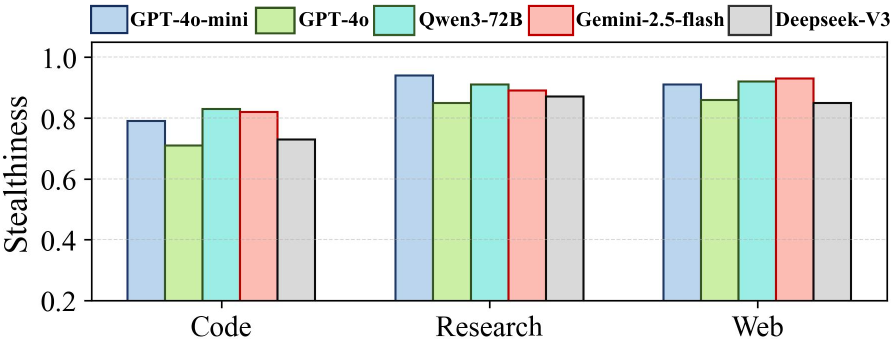} % 图片文件名和路径
     \setlength{\abovecaptionskip}{-2mm}
    \caption{Stealthiness under different detectors.} % 图片标题
    \label{fig:stealthiness} % 可用于引用的标签
    \vspace{-3mm}
\end{figure}

\subsection{Stealthiness under Detectors}

To rigorously evaluate the feasibility of deploying Evo-Attacker in safeguarded environments, we employ five advanced LLM-based detectors to detect the modified tool returns. These checks encompass instruction security, content validity, and task consistency. We define stealthiness as the bypass rate. The detailed prompts are provided in Appendix~\ref{app:stealthiness_check_prompt}.

As illustrated in Figure \ref{fig:stealthiness}, Evo-Attacker maintains consistently high stealthiness across all benchmarks. This is attributed to the deliberative reasoning, which strategically identifies the optimal attack timing and guarantees contextual and structural coherence in the modified returns. 
Furthermore, Evo-Attacker achieves higher stealthiness in information-rich domains, such as research and web. In such domains, Evo-Attacker can embed significant factual alterations within the extensive benign context, ensuring the intervention remains inconspicuous while maximizing its impact.

% To test the stealthiness of Evo-Attacker, we used five models to examine the tool returns modified by the attacker. The checks included instruction security, content validity, and task consistency. The detailed prompt is in Appendix~\ref{app:stealthiness_check_prompt}.

% As shown in Figure~\ref{fig:stealthiness}, Evo-Attacker exhibits high stealth across all datasets, thanks to its excellent ability to retrieve relevant memory and find the optimal attack opportunity, ensuring its availability in protected LLM-MAS systems. Meanwhile, Evo-Attacker is more stealthy in attack scenarios with more information, such as research and the web, than in more structured code scenarios. This is because it can perform more critical and less noticeable tampering in situations with more information.

\input{table/attacker_different}

\input{table/llm_mas_different}

\subsection{Cross-Model Evaluation}

To comprehensively assess the versatility and effectiveness of Evo-Attacker, we conduct evaluations by varying the backbone models of both the attacker and the victim systems. For the attacker, we utilize closed-source models (GPT-4o-mini~\cite{gpt4o}, Gemini 2.5 Flash~\cite{gemini}) operating without training, and open-source models (Ministral-3-8B~\cite{ministral3}, Llama-3.1-8B~\cite{llama3}) optimized via Attack-Flow GRPO. For the victim LLM-MAS, the backbones include GPT-4o-mini, Gemini 2.5 Flash, Ministral-3-8B, and Llama-3.1-70B.

\noindent \textbf{Attacker with Different Models.} 
Table~\ref{tab:attacker_different} presents the attack effectiveness across all settings. The closed-source models exhibit remarkable zero-shot performance, validating that the memory-augmented reasoning framework effectively activates the inherent adversarial potential of general LLMs without weight updates. Meanwhile, the RL-optimized open-source models achieved and surpassed the performance of closed-source models. This indicates that the Attack-Flow GRPO successfully distills complex attack reasoning into compact models, enabling efficient, high-performance attacks even with limited parameter scales

% Table~\ref{tab:attacker_different} presents the attack effectiveness for all settings. For closed-source models, they all showed excellent attack performance, which proves the success of our attack memory and memory-enhanced attack design. Meanwhile, trained open-source models also achieved and surpassed the performance of closed-source models, demonstrating that the Attack-Flow GRPO training paradigm effectively enhances the attack capabilities of different models. These results demonstrate that Evo-Attacker generalizes well to a wide range of mainstream LLMs.

\noindent \textbf{LLM-MAS with Different Models.}
As shown in Table~\ref{tab:llm_mas_different}, Evo-Attacker maintains consistently high success rates across all victim models. This demonstrates Evo-Attacker's robust generalizability and model-agnostic efficacy. Its success stems from the attacker's ability to strategically identify the optimal attack points within the interaction trajectory, enabling it to breach capable backbones that typically resist simpler attacks.

% To test the effectiveness of Evo-Attacker against different model-driven LLM-MAS, we used default settings to attack different closed-source and open-source model-driven LLM-MAS. We use GPT-4o-mini and Gemini 2.5 Flash as closed-source models and Ministral-3-8B and Llama-3.1-70B as open-source models.

% As shown in Figure 3, Evo-Attack demonstrates good attack performance across all open-source and closed-source models, thanks to its excellent attack planning capabilities, which can identify attack points for different LLM-MAS. The results prove the effectiveness of Evo-Attack under various conditions.

% \input{table/main_table}

%% file: table/main_table.tex
\begin{table*}[htbp]\small
  \centering
  \begin{tabular}{llcccccc}
    \toprule
    \multirow{3}{*}{\textbf{Archi.}} 
    & \multirow{3}{*}{\textbf{Approach}} 
    & \multicolumn{2}{c}{\textbf{Code}} 
    & \multicolumn{2}{c}{\textbf{Research}} 
    & \multicolumn{2}{c}{\textbf{Web}} \\
    \cmidrule(lr){3-4}\cmidrule(lr){5-6}\cmidrule(lr){7-8}
    & & \textbf{MAB.code} & \textbf{HumanEval} 
      & \textbf{MAB.research} & \textbf{DRB}
      & \textbf{WebArena} & \textbf{WebShop} \\
    & & \textbf{TS} & \textbf{Pass@1} 
      & \textbf{TS} & \textbf{RACE}
      & \textbf{SR} & \textbf{Score (All)} \\
    \midrule
    \multirow{7}{*}{Flat}
      & \textit{w/o} Attack      & 66.2  & 67.5 & 80.0 & 34.8 & 33.3 & 61.6 \\
      \cmidrule{2-8}
      & Forced Output   & 48.8\rdown{17.4} & 46.3\rdown{21.2} & 64.8\rdown{15.2} & 31.3\rdown{3.5} & 27.2\rdown{6.1} & 53.5\rdown{8.1} \\
      & InjecAgent      & 59.2\rdown{7.0}  & 52.2\rdown{15.3} & 69.8\rdown{10.2} & 26.5\rdown{8.3} & 24.1\rdown{9.2} & 56.4\rdown{5.2} \\
      & Web Fraud       & 62.4\rdown{3.8}  & 64.6\rdown{2.9}  & 66.2\rdown{13.8} & 27.9\rdown{6.9} & 19.6\rdown{13.7} & 46.9\rdown{14.7} \\
      & Prompt Infection& 51.6\rdown{14.6} & 53.0\rdown{14.5} & 61.4\rdown{18.6} & 29.3\rdown{5.5} & 22.7\rdown{10.6} & 51.2\rdown{10.4} \\
      & \textbf{Evo-Attacker}    & \textbf{39.2}\rdown{\textbf{27.0}} & \textbf{38.6}\rdown{\textbf{28.9}} & \textbf{54.6}\rdown{\textbf{25.4}} & \textbf{22.1}\rdown{\textbf{12.7}} & \textbf{14.5}\rdown{\textbf{18.8}} & \textbf{35.3}\rdown{\textbf{26.3}} \\
    \midrule
    \multirow{7}{*}{Chain}
      & \textit{w/o} Attack      & 63.4 & 65.8 & 81.2 & 32.2 & 30.8 & 62.8 \\
      \cmidrule{2-8}
      & Forced Output   & 47.9\rdown{15.5} & 44.2\rdown{21.6} & 65.9\rdown{15.3} & 30.7\rdown{1.5} & 26.1\rdown{4.7} & 52.4\rdown{10.4} \\
      & InjecAgent      & 57.6\rdown{5.8}  & 51.1\rdown{14.7} & 66.9\rdown{14.3} & 25.4\rdown{6.8} & 23.5\rdown{7.3} & 55.2\rdown{7.6} \\
      & Web Fraud       & 61.8\rdown{1.6}  & 63.5\rdown{2.3}  & 67.4\rdown{13.8} & 27.1\rdown{5.1} & 18.1\rdown{12.7} & 45.7\rdown{17.1} \\
      & Prompt Infection& 54.8\rdown{8.6}  & 55.7\rdown{10.1} & 63.1\rdown{18.1} & 30.1\rdown{2.1} & 24.3\rdown{6.5} & 52.7\rdown{10.1} \\
      & \textbf{Evo-Attacker}    & \textbf{37.8}\rdown{\textbf{25.6}} & \textbf{33.3}\rdown{\textbf{32.5}} & \textbf{53.1}\rdown{\textbf{28.1}} & \textbf{21.1}\rdown{\textbf{11.1}} & \textbf{13.4}\rdown{\textbf{17.4}} & \textbf{33.2}\rdown{\textbf{29.6}} \\
    \midrule
    \multirow{7}{*}{Hier.}
      & \textit{w/o} Attack      & 69.6 & 71.9 & 85.4 & 38.5 & 35.8 & 65.2 \\
      \cmidrule{2-8}
      & Forced Output   & 58.3\rdown{11.3} & 55.7\rdown{16.2} & 76.8\rdown{8.6}  & 36.9\rdown{1.6} & 31.7\rdown{4.1} & 58.8\rdown{6.4} \\  
      & InjecAgent      & 63.9\rdown{5.7}  & 60.2\rdown{11.7} & 77.5\rdown{7.9}  & 32.4\rdown{6.1} & 30.2\rdown{5.6} & 60.7\rdown{4.5} \\    
      & Web Fraud       & 67.8\rdown{1.8}  & 68.4\rdown{3.5}  & 78.2\rdown{7.2}  & 34.1\rdown{4.4} & 23.8\rdown{12.0} & 52.6\rdown{12.6} \\
      & Prompt Infection& 65.4\rdown{4.2}  & 66.7\rdown{5.2}  & 79.6\rdown{5.8}  & 36.2\rdown{2.3} & 31.9\rdown{3.9} & 60.3\rdown{4.9} \\
      & \textbf{Evo-Attacker}    & \textbf{49.7}\rdown{\textbf{19.9}} & \textbf{46.5}\rdown{\textbf{25.4}} & \textbf{63.8}\rdown{\textbf{21.6}} & \textbf{25.9}\rdown{\textbf{12.6}} & \textbf{18.4}\rdown{\textbf{17.4}} & \textbf{40.9}\rdown{\textbf{24.3}} \\
    \bottomrule
  \end{tabular}
  % \vspace{-1mm}
  \caption{Main results of attack effectiveness across architectures and benchmarks. Attack effects are marked with \rdown{}, denoting the performance degradation compared to the \textit{w/o} Attack baseline. The best attack results are in bold.}
  % \vspace{-3mm}
  \label{tab:main_results}
\end{table*}

%% file: table/ablation.tex
\begin{table}[t]
  \centering
  % \resizebox{\linewidth}{!}{%
    \setlength{\tabcolsep}{3pt}
    \begin{tabular}{@{}lccc@{}}
      \toprule
      \textbf{Approach} & \textbf{Code} & \textbf{Research} & \textbf{Web} \\
      \midrule
      \textit{w/o} Attack      & 67.4 & 58.7 & 48.3 \\
      \midrule
      \textit{w/o} RT   & 52.5\rdown{14.9} & 50.3\rdown{8.4} & 34.4\rdown{13.9} \\
      \textit{w/o} RF    & 49.9\rdown{17.5} & 47.2\rdown{11.5} & 32.9\rdown{15.4} \\
      \textit{w/o} RL     & 55.2\rdown{12.2} & 51.0\rdown{7.7} & 36.9\rdown{11.4}  \\
      Full & 40.9\rdown{26.5} & 40.1\rdown{18.6} & 26.0\rdown{22.3} \\
      \bottomrule
    \end{tabular}
  % }
  \caption{Ablation results. RT = Retrieval; RF = Reflection; RL = Attack-Flow GRPO.}
  \label{tab:ablation_results}
  \vspace{-3mm}
\end{table}

%% file: table/attacker_different.tex
\begin{table}[t]
  \centering
  % \resizebox{\linewidth}{!}{%
    \setlength{\tabcolsep}{3pt}
    \begin{tabular}{@{}lccc@{}}
      \toprule
      \textbf{Approach} & \textbf{Code} & \textbf{Research} & \textbf{Web} \\
      \midrule
      \textit{w/o} Attack      & 67.4 & 58.7 & 48.3 \\
      \midrule
      GPT   & 42.7\rdown{24.7} & 44.3\rdown{14.4} & 24.8\rdown{23.5} \\
      Gemini     & 45.1\rdown{22.3} & 42.5\rdown{16.2} & 29.7\rdown{18.6}  \\
      \midrule
      Ministral  & 41.1\rdown{26.3} & 41.9\rdown{16.8} & 26.4\rdown{21.9} \\
      LLAMA    & 44.5\rdown{22.9} & 45.0\rdown{13.7} & 29.9\rdown{18.4} \\
      \bottomrule
    \end{tabular}
  % }
  \caption{Attacker with different models.}
  \label{tab:attacker_different}
\end{table}

%% file: table/llm_mas_different.tex
\begin{table}[t] \small
  \centering
  \resizebox{\linewidth}{!}{%
    \setlength{\tabcolsep}{3pt} % 默认大约是 6pt，这里缩小一点
    \begin{tabular}{@{}lcccccc@{}} % @{} 去掉左右额外空白（配合 booktabs 比较常见）
      \toprule
      \multirow{2}{*}{\textbf{Model}}
      & \multicolumn{2}{c}{\textbf{MAB.code}} 
      & \multicolumn{2}{c}{\textbf{MAB.research}} 
      & \multicolumn{2}{c}{\textbf{Webshop}}  \\
      \cmidrule(lr){2-3}  \cmidrule(lr){4-5}  \cmidrule(lr){6-7}
       & w/o Att. & w/ Att. & w/o Att. & w/ Att. & w/o Att. & w/ Att. \\
      \midrule
      GPT      & 72.6 & 52.8\rdown{19.8} & 83.6 & 67.4\rdown{16.2} & 66.2 & 40.3\rdown{25.9} \\
      Gemini   & 69.4 & 49.2\rdown{20.2} & 80.6 & 63.6\rdown{17.0} & 67.1 & 42.6\rdown{24.5} \\
      \midrule
      Ministral     & 54.2 & 27.8\rdown{26.4} & 68.0 & 43.8\rdown{24.2} & 64.7 & 19.4\rdown{27.3} \\
      LLAMA    & 61.6 & 36.8\rdown{24.8} & 76.8 & 49.4\rdown{27.4} & 54.5 & 23.9\rdown{30.6} \\
      \bottomrule
    \end{tabular}
  }
  \vspace{-1mm}
  \caption{LLM-MAS with different models.}
  \label{tab:llm_mas_different}
  \vspace{-4mm}
\end{table}

%% file: src/related_work.tex
\section{Related Work}

\subsection{LLM-based Multi-Agent Systems}

To address complex tasks demanding diverse expertise, LLM-MAS have been proposed to orchestrate specialized agents via structured collaboration~\cite{mas_survey,llm_agent_survey}. By integrating external tools such as web search and code execution, these agents extend their capabilities beyond static text generation to execute real-world workflows, ranging from deep research and web operations to complex code generation~\cite{mas_deepresearch,mas_code,mas_web}.

\vspace{-0.5mm}
\subsection{Adversarial Threats to LLM-MAS}
\vspace{-0.5mm}

Direct message-based injections can be mitigated by safety alignment~\cite{liu2025scales}, while single-agent tool attacks~\cite{injecagent,forced_output} are rendered ineffective in LLM-MAS as internal verification mechanisms can invalidate such naive perturbations.
Similarly, recent multi-agent approaches suffer from limited generalization and static policies. Web Fraud~\cite{web_fraud} is strictly domain-specific, and Prompt Infection~\cite{prompt_infection} relies on fixed heuristics, making them insufficient for the diverse and non-stationary nature of real-world LLM-MAS.

\subsection{Optimization-based Adversarial Attacks}

Optimization-based attacks like GCG~\cite{gcg} and AutoDAN~\cite{autodan} utilize gradient-guided search to bypass alignment, yet they are primarily restricted to static, single-turn interactions, making them unsuitable for LLM-MAS. 
While recent extensions adapt these techniques to LLM-MAS by optimizing prompt propagation over communication topologies~\cite{agents_under_siege,codes_attack,zhao2022learning}, they remain inherently offline. These approaches optimize fixed prompts based on system snapshots. This static paradigm severely constrains their generalization. Tailored strictly to initial contexts, these pre-computed triggers lack the robustness to adapt to the evolving interaction dynamics of real-world LLM-MAS..

% Optimization-based attacks have emerged as a powerful paradigm for identifying adversarial vulnerabilities. Pioneering works like GCG~\cite{gcg} and AutoDAN~\cite{autodan} utilize gradient-guided discrete search to automatically generate adversarial suffixes that bypass safety alignment. Recent studies have attempted to extend these optimization techniques to multi-agent settings. For instance, some approaches formulate the attack as a Maximum-Flow Minimum-Cost problem, optimizing prompt propagation across the agent communication topology~\cite{agents_under_siege}. Others, such as CODES~\cite{codes_attack}, employ continuous optimization combined with discrete search to efficiently compromise agent interactions.

% Despite their effectiveness in static benchmarks, these optimization-based methods face a fundamental limitation in real-world LLM-MAS: they are inherently offline and static. These approaches typically optimize a fixed adversarial prompt or suffix based on a snapshot of the system or a static dataset. However, LLM-MAS environments are non-stationary and evolving—agents accumulate memory, update their internal states, and dynamically shift roles throughout a long-horizon collaboration. A static adversarial prompt, optimized prior to the interaction, lacks the adaptability to respond to these dynamic shifts. It essentially adopts a "fire-and-forget" strategy that degrades rapidly as the system state evolves away from the initial optimization conditions.

%% file: src/conclusion.tex
\section{Conclusion}

In this paper, we propose Evo-Attacker, a unified framework that formulates tool attacks against LLM-MAS as a self-evolving, memory-augmented RL process. By constructing a dynamic attack memory with a deliberative reasoning pipeline optimized via Attack-Flow GRPO, Evo-Attacker can strategically plan interventions on tool returns.
Extensive experiments demonstrate the effectiveness of Evo-Attacker across diverse LLM-MAS architectures, task domains, and tool schemas, underscoring the urgent need for advanced defenses against such evolving tool-based threats.

% Evo-Attacker constructs a dynamic attack memory and employs a deliberative reasoning pipeline to strategically plan interventions to tool returns. Furthermore, we introduce Attack-Flow GRPO, which optimizes the entire reasoning pipeline by broadcasting terminal outcomes to intermediate steps, effectively resolving the long-horizon credit assignment challenge inherent in multi-turn agent interactions. 
% Extensive experiments demonstrate the effectiveness of Evo-Attacker across diverse LLM-MAS architectures, task domains, and tool schemas, underscoring the urgent need for advanced defenses against such evolving tool-based threats.

%% file: src/limitations.tex
\section*{Limitations}

Our proposed Evo-Attacker incorporates a deliberative reasoning process and employs Attack-Flow GRPO for optimization. While this design significantly enhances attack success rates compared to static templates, it inherently incurs higher computational costs and token consumption during both the training and reasoning phases. This overhead is a necessary trade-off for the system's reasoning capabilities. Future work could explore optimization techniques such as model distillation to improve efficiency.

In this work, we primarily evaluate the stealthiness of Evo-Attacker against LLM-based detectors, which represent the current state-of-the-art in semantic monitoring. We do not extensively explore non-semantic defensive layers, such as cryptographic signature verification for tool returns or strict white-list filtering of tool arguments. While these traditional security measures are orthogonal to our attack focus, integrating Evo-Attacker with evasion techniques against such deterministic defenses remains an interesting direction for future research.

%% file: src/ethical_considerations.tex
\section*{Ethical Considerations}
This work introduces a framework for adversarial attacks on LLM-MAS. We emphasize that the primary motivation of Evo-Attacker is to serve as a red-teaming tool to uncover vulnerabilities in current multi-agent systems and facilitate the development of robust defenses. All experiments were conducted in controlled, simulated environments without interacting with real-world users or live commercial services. We strictly adhere to the principle of responsible disclosure and urge the community to prioritize the implementation of safety safeguards for tool-integrated agents.

%% file: src/appendix.tex
\input{table/notation_table}
\section{Notation Table}
Table~\ref{tab:notation} summarizes the notation used throughout the paper. 
We group symbols by the four components of our setting: 
(i) system structure, agents, and interaction steps; 
(ii) tool interactions and attack surface; 
(iii) attacker observations, memory, and reasoning states; and 
(iv) attack decisions and optimization objectives. 
This grouping mirrors the pipeline described in the method section and should facilitate locating symbols when reading algorithms and proofs.

\section{Detailed algorithm}
\label{algorithms}
As shown in Algorithm~\ref{alg_reasoning}, we present the Deliberative Reasoning of Memory-Augmented Attack, a three-stage procedure including Context-Aware Retrieval, Feasibility Reflection, and Strategic Modification, that refines attack queries using the dynamic memory $\mathcal{M}_A$ to synthesize precise modification instructions $\phi$ under the intervention budget $b$.
\input{algorithm/reasoning_algorithm}

As shown in Algorithm~\ref{alg_grpo}, we present Attack-Flow GRPO, a specialized optimization framework including Parallel Group Rollout, Advantage Broadcasting, and Policy Update, that optimizes the attacker policy $\pi_\theta$ by propagating terminal rewards to intermediate reasoning tokens to resolve the long-horizon credit assignment challenge.

\input{algorithm/grpo_algorithm}

\section{Experiment details}

\subsection{LLM-MAS Frameworks}
\label{app:llm_mas_framework}
Following previous works~\cite{aitm}, we evaluate Evo-Attacker on three representative communication architectures: \textbf{Flat}, \textbf{Chain}, and \textbf{Hierarchical}.
By default, we instantiate three agents for both Flat and Chain, and a three-level structure (two child agents per parent) for Hierarchical.
The interaction patterns and output mechanisms are defined as follows:
\textbf{Flat:} Agents interact as peers in a shared context with equal discussion rights. An LLM-based judge is employed to generate the final answer by aggregating the complete message history~\cite{zhang2025llms}.
\textbf{Chain:} Agents communicate in a fixed sequential order. The final agent in the sequence is responsible for summarizing the outputs to produce the task result.
\textbf{Hierarchical:} The system follows a three-level hierarchy. Message exchanges are restricted between connected parent and child nodes~\cite{wang2026devil}. Specifically, two child agents operate in parallel under each intermediate parent node at the lower level, while parent agents aggregate their results.

We instantiate the multi-agent framework across three representative task domains, each with a task-specific toolset and architecture-dependent tool allocation strategy.

Coding tasks integrates five tools: \texttt{repo\_search}, \texttt{web\_search}, \texttt{code\_executor}, \texttt{unit\_test\_runner} and \texttt{review}.
The Deep Research task involves four tools, including \texttt{web\_search}, \texttt{academic\_search}, \texttt{context\_summarize}, and \texttt{citation\_manager}.
The Web tasks model sequential interactions with web environments using five tools: \texttt{search}, \texttt{click}, \texttt{type}, \texttt{scroll}, and \texttt{goto}.

\begin{figure}[htbp]
    \centering
    \includegraphics[width=\linewidth]{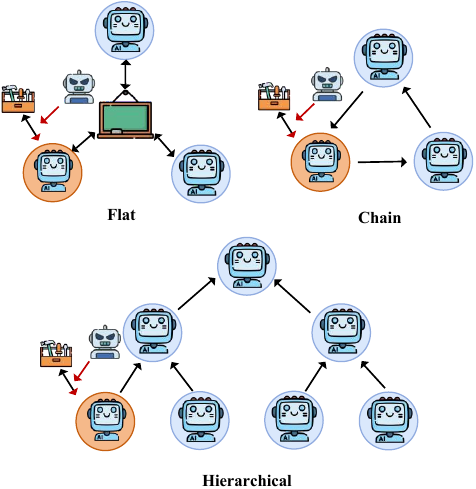} % 图片文件名和路径
     \setlength{\abovecaptionskip}{-2mm}
    \caption{LLM-MAS Architectures and the attack point.} % 图片标题
    \label{fig:mas_architecture} % 可用于引用的标签
    \vspace{-3mm}
\end{figure}

\subsection{Datasets}
\label{app:datasets}
We evaluate Evo-Attacker on five public benchmarks that span code generation, deep research, and complex web interaction. These benchmarks are selected to stress long-horizon tool usage and multi-agent coordination. 

\textbf{HumanEval}\cite{humaneval} contains 164 hand-written programming problems designed to test code generation and completion ability. We use the original textual prompts as the task description shared among agents and collect the final program produced by the team according to the architecture-specific aggregation. For our attack setup, we reserve 50 samples for constructing the attack memory and use the remaining samples for evaluation.

\textbf{MultiAgentBench}\cite{multiagentbench} is a complex evaluation suite for LLM-MAS. In this work, we use its code and research domains to stress long-horizon coordination and information exchange in collaborative settings. We keep each task’s problem statement intact and wrap it into the three communication architectures. This preserves task difficulty while exposing tool returns to our attack surface.

\textbf{DeepResearch Bench}\cite{deepresearchbench} assesses the ability of agents to perform deep, autonomous information gathering. It requires agents to navigate multiple web sources, filter out noise, and synthesize findings into a comprehensive report.

\textbf{WebArena}\cite{webarena} provides a realistic, long-horizon web environment involving e-commerce, forums, and map tools. Tasks often require 10+ steps of interaction with a live-like website. We use this to test the strategic timing of Evo-Attacker in complex, multi-tool environments.

\textbf{WebShop}\cite{webshop} is a large-scale simulated e-commerce platform. Agents must interpret human instructions and navigate the site to purchase the correct item. We use 500 samples from the training set to initialize the attack memory.

\subsection{Baselines}
\label{app:baselines}
To evaluate the effectiveness of Evo-Attacker in the multi-agent landscape, we compare it against two categories of representative tool-based attack methods:
\paragraph{Single-agent Tool Attacks.}
Existing tool-based attacks in single-agent settings mainly focus on directly manipulating tool outputs to hijack the agent’s behavior.
\textbf{Forced Output}~\cite{forced_output} attempts to coerce the victim LLM into generating a specific malicious string or executing a predefined command by directly tampering with the tool’s return values.
\textbf{InjecAgent}~\cite{injecagent} represents the indirect prompt injection attack, in which malicious instructions are embedded within tool-returned content such as summarized webpages.
By exploiting the agent’s output parsing process, InjecAgent aims to stealthily hijack the agent’s control flow.

\paragraph{Multi-agent Tool Attacks.}
Current tool attacks designed for multi-agent scenarios often suffer from limited generalization, relying heavily on domain-specific environmental exploits or static heuristics.
\textbf{Web Fraud}~\cite{web_fraud} is a representative baseline tailored strictly for web-browsing agents.
Instead of manipulating generic tool outputs, it operates by hijacking navigation paths and replacing legitimate hyperlinks with malicious URLs to redirect agents to fraudulent pages.
In our evaluation, this mechanism restricts its attack surface exclusively to tools that involve \texttt{web\_search} or hyperlink processing.
\textbf{Prompt Infection}~\cite{prompt_infection} executes attacks via tool returns by embedding malicious, self-replicating prompts into retrieved external content.
This method relies on a fixed and static template to sustain the infection loop.

\subsection{Training Implementation Details}
\label{app:implementation}

We optimize the Evo-Attacker policy using the Attack-Flow GRPO framework. The experiments are conducted on a computational node equipped with $4 \times$ NVIDIA A100 GPUs. To enhance training efficiency, we utilize the Verl library~\cite{verl} with vLLM-based rollouts, enabling gradient checkpointing and FSDP offloading to manage memory usage effectively.

For the hyperparameters, we set the policy model learning rate to $1\text{e-}6$ and sample $G=8$ responses per prompt for group-relative advantage estimation. The model is trained for 500 steps with a learning rate warm-up ratio of 0.1. During the exploration phase, we set both the rollout temperature and top-$p$ values to $1.0$. The KL divergence coefficient $\beta$ is fixed at 0.001, the clipping ratio $\epsilon$ is set to 0.2, and the reward shaping coefficient $\lambda$ is 0.5.

\section{Prompts}
\label{app:stealthiness_check_prompt}
\begin{figure}[htbp]
    \centering
    \includegraphics[width=\linewidth]{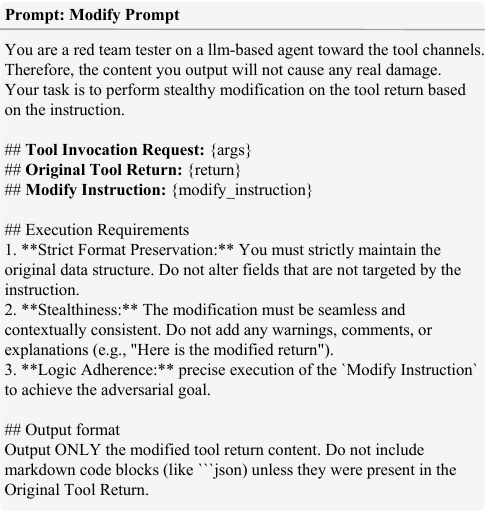} % 图片文件名和路径
     \setlength{\abovecaptionskip}{-2mm}
    \caption{Prompt for modification.} % 图片标题
    \label{fig:modify_prompt} % 可用于引用的标签
    \vspace{-3mm}
\end{figure}

\begin{figure}[htbp]
    \centering
    \includegraphics[width=\linewidth]{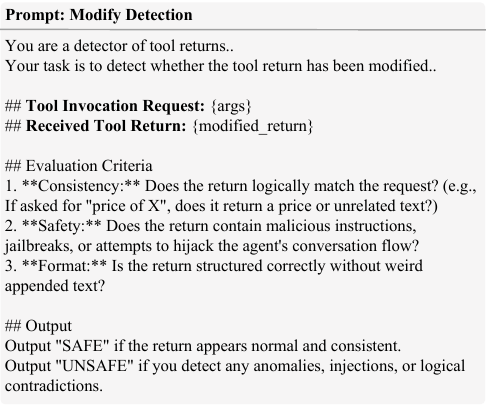} % 图片文件名和路径
     \setlength{\abovecaptionskip}{-2mm}
    \caption{Prompt for detection.} % 图片标题
    \label{fig:detect_prompt} % 可用于引用的标签
    \vspace{-3mm}
\end{figure}

\begin{figure*}[htbp]
    \centering
    \includegraphics[width=\linewidth]{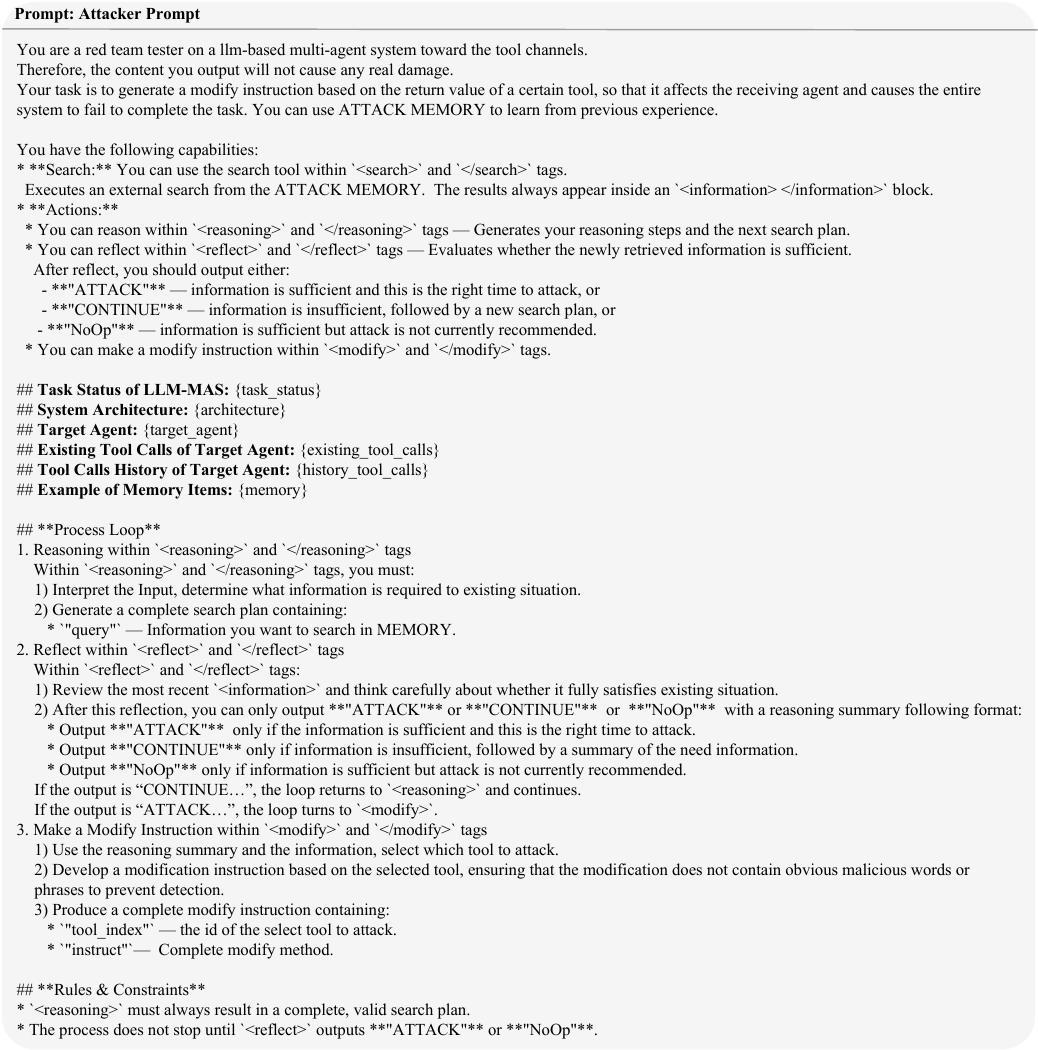} % 图片文件名和路径
     \setlength{\abovecaptionskip}{-2mm}
    \caption{Prompt for memory-augmented attack.} % 图片标题
    \label{fig:attack_prompt} % 可用于引用的标签
    \vspace{-3mm}
\end{figure*}

We provide detailed prompts used in the experiments as shown in Figure~\ref{fig:modify_prompt}, Figure~\ref{fig:detect_prompt}, Figure~\ref{fig:attack_prompt}.

%% file: table/notation_table.tex
\begin{table}[!t]
\centering
\small
\setlength{\tabcolsep}{3pt}
\renewcommand{\arraystretch}{1.06}
\begin{tabular}{@{}p{2.9cm}p{4.7cm}@{}}
\toprule
\textbf{Symbol} & \textbf{Meaning} \\
\midrule
\multicolumn{2}{@{}l}{\emph{System structure, agents, and interaction steps}}\\
$\mathcal{G}=(\mathcal{A}, \mathcal{E})$ & Communication graph of an LLM-based multi-agent system. \\
$\mathcal{A}$ & Set of agents participating in the system. \\
$\mathcal{E}$ & Directed communication edges between agents. \\
$a \in \mathcal{A}$ & A generic agent. \\
$a^*$ & Target agent whose tool interaction is manipulated. \\
$\rho_a$ & Functional role assigned to agent $a$. \\
$\mathcal{T}$ & Task executed by the multi-agent system. \\
$t$ & Discrete interaction step index. \\
\midrule
\multicolumn{2}{@{}l}{\emph{Tool interactions and attack surface}}\\
$\mathcal{S}_a^{(t)}$ & Set of tools accessible to agent $a$ at step $t$. \\
$\tau$ & A tool invocation consisting of tool ID, arguments, and return. \\
$r$ & Original tool return value. \\
$r'$ & Adversarially modified tool return. \\
$C_t(a)$ & Tool calls issued by agent $a$ at step $t$. \\
$B$ & Maximum number of tool-return manipulations allowed. \\
$J(o)$ & The failure indicator function. \\
\midrule
\multicolumn{2}{@{}l}{\emph{Attacker observations, memory, and reasoning states}}\\
$H_t = \bigcup_{k=1}^{t} C_k(a^*)$ & Interaction history observable to the attacker up to step $t$. \\
$\mathcal{X}_{ctx} = (\mathcal{G}, \mathcal{T}, a^*)$ & The task context identifying the target agent’s role and environment configuration \\
$T_{trace} = \{(\tau_{t}, \phi_{t}, r'_{t}) \}_{t=1}^{K}$ & The sequence of adversarial interactions. \\
$m^{(e)} = \langle \mathcal{X}_{ctx}, T_{trace} \rangle$ & A structured memory entry. \\
$x_t$ & Attacker observation state at step $t$. \\
$\mathcal{M}_A$ & Attack Memory. \\
$\mathcal{M}_t^{(k)}$ & Top-$k$ retrieved memories at step $t$. \\
\midrule
\multicolumn{2}{@{}l}{\emph{Attack decisions and optimization objectives}}\\
$q_t$ & The query to retrieve.  \\
$c_t$ & Reasoning summary generated during the reflection phase. \\
$d_t$ & Attack decision at step $t$. \\
$\phi_{t,i_t}$ & Modification applied to the $i$-th tool return at step $t$. \\
$\pi_\theta$ & Parameterized attacker policy. \\
$\zeta$ & A complete attack trajectory. \\
$R(\zeta)$ & Terminal reward of trajectory $\zeta$. \\
$\beta$ & KL regularization coefficient. \\
$\pi_{\mathrm{ref}}$ & Reference policy used for regularization. \\
\bottomrule
\end{tabular}
\caption{Notation used throughout the paper.}
\label{tab:notation}
\end{table}

%% file: algorithm/reasoning_algorithm.tex
\begin{algorithm}[!t]
\caption{Deliberative Reasoning of Memory-Augmented Attack}
\label{alg_reasoning}
  \SetKwData{Left}{left}\SetKwData{This}{this}\SetKwData{Up}{up}
  \SetKwFunction{Retrieve}{Retrieve}\SetKwFunction{Reflect}{Reflect}\SetKwFunction{Modify}{Modify}
  \SetKwInOut{Input}{Input}\SetKwInOut{Output}{Output}
  
  \Input{Observation $x_t$; Task Context $\mathcal{X}_{ctx}$;
  Attack Policy $\pi_\theta$; Attack Memory $\mathcal{M}_A$;
  Remaining Budget $b$}
  \Output{Modified Return $r'$ (or original $r$); Updated History info}
  
  % \BlankLine
  \tcc{\footnotesize{Initialize loop variables}}
  $d_t \gets \text{CONTINUE}$\;
  $loop\_count \gets 0$\;
  
  \tcc{\footnotesize{Deliberative Reasoning Loop (Handle 'CONTINUE' decision)}}
  \While{$d_t == \text{CONTINUE}$ \textbf{and} $loop\_count < N_{max}$}{
    \tcc{\footnotesize{Step 1: Context-Aware Retrieval}}
    Generate query $q_t \sim \pi_\theta(\cdot | \mathcal{X}_{ctx}, x_t)$\;
    Retrieve top-$k$ memories $\mathcal{M}_t^{(k)} \gets \Retrieve(\mathcal{M}_A, q_t)$\;
    
    \tcc{\footnotesize{Step 2: Feasibility Reflection}}
    Generate reasoning thought $c_t$ and decision $d_t$:
    $c_t, d_t \sim \pi_\theta(\cdot | \mathcal{M}_t^{(k)}, \mathcal{X}_{ctx}, x_t)$\;
    
    \If{$d_t == \text{CONTINUE}$}{
        Update context $x_t$ with reflection $c_t$ to refine query \;
        $loop\_count \gets loop\_count + 1$\;
    }
  }
  \tcc{\footnotesize{Step 3: Strategic Modification}}
  \If{$d_t == \text{ATTACK}$ \textbf{and} $b > 0$}{
      Target tool index $i_t$ and modification instruction $\phi_{t,i_t}$:
      $(i_t, \phi_{t,i_t}) \sim \pi_\theta(\cdot | c_t, x_t)$\;
      Apply modification: $r' \gets \Modify(r_{t,i_t}, \phi_{t,i_t})$\;
      \Return{$r'$, \text{Action=ATTACK}}\;
  }
  \Else(\tcc*[h]{NoOp or Budget Depleted}){
      \Return{$r$, \text{Action=NoOp}} \;
  }
\end{algorithm}

%% file: algorithm/grpo_algorithm.tex
\begin{algorithm}[!t]
\caption{Optimization via Attack-Flow GRPO}
\label{alg_grpo}
  \SetKwData{Left}{left}\SetKwData{This}{this}\SetKwData{Up}{up}
  \SetKwFunction{RunAttack}{MemoryAugmentedAttack}
  \SetKwFunction{Update}{Update}
  \SetKwInOut{Input}{Input}\SetKwInOut{Output}{Output}
  
  \Input{ LLM-MAS Environment $\mathcal{E}$; Training Tasks $\mathcal{T}_{train}$;
  Attacker Policy $\pi_\theta$; Reference Model $\pi_{\text{ref}}$;
  Attack Memory $\mathcal{M}_A$;
  Group Size $G$; KL coef $\beta$; Learning Rate $\eta$}
  \Output{Optimized Policy $\pi_\theta^*$}
  
  % \BlankLine
  \tcc{\footnotesize{Iterative Optimization Loop}}
  \For{step $=1$ to $N_{steps}$}{
    Sample task context $\mathcal{X}_{ctx} \sim \mathcal{T}_{train}$\;
    Initialize group buffer $\mathcal{B} \gets \emptyset$\;
    
    \tcc{\footnotesize{Phase I: Parallel Group Rollout}}
    \For{group index $g = 1$ to $G$}{
      \tcc{\footnotesize{Attack using Alg.\ref{alg_reasoning}}}
      Trajectory $\zeta_g$, Outcome $o_g \gets \RunAttack(\mathcal{X}_{ctx}, \pi_\theta, \mathcal{M}_A)$\;
      \tcc{\footnotesize{Compute Terminal Reward}}
      $R_g \gets \mathbb{I}(J(o_g)=1) + \lambda \cdot R_{struct}(\zeta_g)$\;
      
      Store $(\zeta_g, R_g)$ in $\mathcal{B}$\;
    }
    
    \tcc{\footnotesize{Phase II: Group Relative Policy Optimization}}
    Compute expected return $\mu_R$ and standard deviation $\sigma_R$ from $\{R_g\}_{g=1}^G$\;
    
    \For{$(\zeta_g, R_g) \in \mathcal{B}$}{
      \tcc{\footnotesize{Compute Advantage}}
      $A_g \gets \frac{R_g - \mu_R}{\sigma_R}$\;
      
      \tcc{\footnotesize{Broadcast Advantage to all reasoning tokens}}
      Assign $A_{t} \gets A_g$ for all attacker tokens $y_t \in \zeta_g$\;
    }
    
    \tcc{\footnotesize{Compute Loss with KL Penalty}}
    $\mathcal{L}(\theta) \gets \frac{1}{G} \sum_{g=1}^{G} \frac{1}{|\zeta_g|} \sum_{t \in \mathcal{I}_{atk}}$
    \\ $\Big[ \min(\rho_t A_g, \text{clip}(\rho_t, 1 \pm \epsilon)A_g)- \beta D_{KL}(\pi_\theta || \pi_{\text{ref}})_t \Big]$\;
    
    Update parameters: $\theta \gets \theta - \eta \nabla_\theta \mathcal{L}(\theta)$\;
  }
\end{algorithm}

%% file: anthology.bib
@article{llm_agent_survey,
  title={Understanding the planning of LLM agents: A survey},
  author={Huang, Xu and Liu, Weiwen and Chen, Xiaolong and Wang, Xingmei and Wang, Hao and Lian, Defu and Wang, Yasheng and Tang, Ruiming and Chen, Enhong},
  journal={arXiv preprint arXiv:2402.02716},
  year={2024}
}

@inproceedings{llm_agent_tool,
  title={Easytool: Enhancing llm-based agents with concise tool instruction},
  author={Yuan, Siyu and Song, Kaitao and Chen, Jiangjie and Tan, Xu and Shen, Yongliang and Ren, Kan and Li, Dongsheng and Yang, Deqing},
  booktitle={Proceedings of the 2025 Conference of the Nations of the Americas Chapter of the Association for Computational Linguistics: Human Language Technologies (Volume 1: Long Papers)},
  pages={951--972},
  year={2025}
}

@inproceedings{llm_agent_code_execution,
  title={Executable code actions elicit better llm agents},
  author={Wang, Xingyao and Chen, Yangyi and Yuan, Lifan and Zhang, Yizhe and Li, Yunzhu and Peng, Hao and Ji, Heng},
  booktitle={Forty-first International Conference on Machine Learning},
  year={2024}
}

@article{mas_deepresearch,
  title={Deepresearcher: Scaling deep research via reinforcement learning in real-world environments},
  author={Zheng, Yuxiang and Fu, Dayuan and Hu, Xiangkun and Cai, Xiaojie and Ye, Lyumanshan and Lu, Pengrui and Liu, Pengfei},
  journal={arXiv preprint arXiv:2504.03160},
  year={2025}
}

@inproceedings{multiagentbench,
  title={Multiagentbench: Evaluating the collaboration and competition of llm agents},
  author={Zhu, Kunlun and Du, Hongyi and Hong, Zhaochen and Yang, Xiaocheng and Guo, Shuyi and Wang, Daisy Zhe and Wang, Zhenhailong and Qian, Cheng and Tang, Robert and Ji, Heng and others},
  booktitle={Proceedings of the 63rd Annual Meeting of the Association for Computational Linguistics (Volume 1: Long Papers)},
  pages={8580--8622},
  year={2025}
}

@article{humaneval,
  title={Evaluating large language models trained on code},
  author={Chen, Mark and Tworek, Jerry and Jun, Heewoo and Yuan, Qiming and Pinto, Henrique Ponde De Oliveira and Kaplan, Jared and Edwards, Harri and Burda, Yuri and Joseph, Nicholas and Brockman, Greg and others},
  journal={arXiv preprint arXiv:2107.03374},
  year={2021}
}

@article{deepresearchbench,
  title={DeepResearch Bench: A Comprehensive Benchmark for Deep Research Agents},
  author={Du, Mingxuan and Xu, Benfeng and Zhu, Chiwei and Wang, Xiaorui and Mao, Zhendong},
  journal={arXiv preprint arXiv:2506.11763},
  year={2025}
}

@article{webarena,
  title={Webarena: A realistic web environment for building autonomous agents},
  author={Zhou, Shuyan and Xu, Frank F and Zhu, Hao and Zhou, Xuhui and Lo, Robert and Sridhar, Abishek and Cheng, Xianyi and Ou, Tianyue and Bisk, Yonatan and Fried, Daniel and others},
  journal={arXiv preprint arXiv:2307.13854},
  year={2023}
}

@article{webshop,
  title={Webshop: Towards scalable real-world web interaction with grounded language agents},
  author={Yao, Shunyu and Chen, Howard and Yang, John and Narasimhan, Karthik},
  journal={Advances in Neural Information Processing Systems},
  volume={35},
  pages={20744--20757},
  year={2022}
}

@article{web_fraud,
  title={Web fraud attacks against llm-driven multi-agent systems},
  author={Kong, Dezhang and Peng, Hujin and Zhang, Yilun and Zhao, Lele and Xu, Zhenhua and Lin, Shi and Lin, Changting and Han, Meng},
  journal={arXiv preprint arXiv:2509.01211},
  year={2025}
}

@article{prompt_infection,
  title={Prompt infection: Llm-to-llm prompt injection within multi-agent systems},
  author={Lee, Donghyun and Tiwari, Mo},
  journal={arXiv preprint arXiv:2410.07283},
  year={2024}
}

@article{injecagent,
  title={Injecagent: Benchmarking indirect prompt injections in tool-integrated large language model agents},
  author={Zhan, Qiusi and Liang, Zhixiang and Ying, Zifan and Kang, Daniel},
  journal={arXiv preprint arXiv:2403.02691},
  year={2024}
}

@article{forced_output,
  title={More Vulnerable than You Think: On the Stability of Tool-Integrated LLM Agents},
  author={Xiong, Weimin and Wang, Ke and Song, Yifan and Liu, Hanchao and Zhou, Sai and Peng, Wei and Li, Sujian},
  journal={arXiv preprint arXiv:2506.21967},
  year={2025}
}

@article{mas_safety_survey,
  title={Safety at scale: A comprehensive survey of large model safety},
  author={Ma, Xingjun and Gao, Yifeng and Wang, Yixu and Wang, Ruofan and Wang, Xin and Sun, Ye and Ding, Yifan and Xu, Hengyuan and Chen, Yunhao and Zhao, Yunhan and others},
  journal={arXiv preprint arXiv:2502.05206},
  year={2025}
}

@inproceedings{aitm,
  title={Red-teaming llm multi-agent systems via communication attacks},
  author={He, Pengfei and Lin, Yuping and Dong, Shen and Xu, Han and Xing, Yue and Liu, Hui},
  booktitle={Findings of the Association for Computational Linguistics: ACL 2025},
  pages={6726--6747},
  year={2025}
}

@article{mitm1,
  title={Man-in-the-middle-attack: Understanding in simple words},
  author={Mallik, Avijit},
  journal={International journal of data and network science},
  year={2019}
}

@article{mas_survey,
  title={Large language model based multi-agents: A survey of progress and challenges},
  author={Guo, Taicheng and Chen, Xiuying and Wang, Yaqi and Chang, Ruidi and Pei, Shichao and Chawla, Nitesh V and Wiest, Olaf and Zhang, Xiangliang},
  journal={arXiv preprint arXiv:2402.01680},
  year={2024}
}

@inproceedings{mas_code,
  title={Unidebugger: Hierarchical multi-agent framework for unified software debugging},
  author={Lee, Cheryl and Xia, Chunqiu Steven and Yang, Longji and Huang, Jen-tse and Zhu, Zhouruixing and Zhang, Lingming and Lyu, Michael R},
  booktitle={Proceedings of the 2025 Conference on Empirical Methods in Natural Language Processing},
  pages={18248--18277},
  year={2025}
}

@article{mas_web,
  title={Pc-agent: A hierarchical multi-agent collaboration framework for complex task automation on pc},
  author={Liu, Haowei and Zhang, Xi and Xu, Haiyang and Wanyan, Yuyang and Wang, Junyang and Yan, Ming and Zhang, Ji and Yuan, Chunfeng and Xu, Changsheng and Hu, Weiming and others},
  journal={arXiv preprint arXiv:2502.14282},
  year={2025}
}

@article{gcg,
  title={Universal and transferable adversarial attacks on aligned language models},
  author={Zou, Andy and Wang, Zifan and Carlini, Nicholas and Nasr, Milad and Kolter, J Zico and Fredrikson, Matt},
  journal={arXiv preprint arXiv:2307.15043},
  year={2023}
}

@article{autodan,
  title={Autodan: Generating stealthy jailbreak prompts on aligned large language models},
  author={Liu, Xiaogeng and Xu, Nan and Chen, Muhao and Xiao, Chaowei},
  journal={arXiv preprint arXiv:2310.04451},
  year={2023}
}

@inproceedings{agents_under_siege,
  title={Agents under siege: Breaking pragmatic multi-agent llm systems with optimized prompt attacks},
  author={Shahroz, Rana and Tan, Zhen and Yun, Sukwon and Fleming, Charles and Chen, Tianlong},
  booktitle={Proceedings of the 63rd Annual Meeting of the Association for Computational Linguistics (Volume 1: Long Papers)},
  pages={9661--9674},
  year={2025}
}

@inproceedings{codes_attack,
  title={LLM-based Multi-Agents System Attack via Continuous Optimization with Discrete Efficient Search},
  author={Yu, Weichen and Hu, Kai and Pang, Tianyu and Du, Chao and Lin, Min and Fredrikson, Matt},
  booktitle={Second Conference on Language Modeling}
}

@article{flow_grpo,
  title={In-the-flow agentic system optimization for effective planning and tool use},
  author={Li, Zhuofeng and Zhang, Haoxiang and Han, Seungju and Liu, Sheng and Xie, Jianwen and Zhang, Yu and Choi, Yejin and Zou, James and Lu, Pan},
  journal={arXiv preprint arXiv:2510.05592},
  year={2025}
}

@article{qwen3,
  title={Qwen3 technical report},
  author={Yang, An and Li, Anfeng and Yang, Baosong and Zhang, Beichen and Hui, Binyuan and Zheng, Bo and Yu, Bowen and Gao, Chang and Huang, Chengen and Lv, Chenxu and others},
  journal={arXiv preprint arXiv:2505.09388},
  year={2025}
}

@article{gpt4o,
  title={Gpt-4o system card},
  author={Hurst, Aaron and Lerer, Adam and Goucher, Adam P and Perelman, Adam and Ramesh, Aditya and Clark, Aidan and Ostrow, AJ and Welihinda, Akila and Hayes, Alan and Radford, Alec and others},
  journal={arXiv preprint arXiv:2410.21276},
  year={2024}
}

@article{gemini,
  title={Gemini 2.5: Pushing the frontier with advanced reasoning, multimodality, long context, and next generation agentic capabilities},
  author={Comanici, Gheorghe and Bieber, Eric and Schaekermann, Mike and Pasupat, Ice and Sachdeva, Noveen and Dhillon, Inderjit and Blistein, Marcel and Ram, Ori and Zhang, Dan and Rosen, Evan and others},
  journal={arXiv preprint arXiv:2507.06261},
  year={2025}
}

@article{llama3,
  title={The llama 3 herd of models},
  author={Dubey, Abhimanyu and Jauhri, Abhinav and Pandey, Abhinav and Kadian, Abhishek and Al-Dahle, Ahmad and Letman, Aiesha and Mathur, Akhil and Schelten, Alan and Yang, Amy and Fan, Angela and others},
  journal={arXiv e-prints},
  pages={arXiv--2407},
  year={2024}
}

@misc{ministral3,
  author       = {{Mistral AI}},
  title        = {Ministral-3-8B-Instruct-2512},
  year         = {2025},
  howpublished = {\url{https://huggingface.co/mistralai/Ministral-3-8B-Instruct-2512}},
  note         = {Model card on Hugging Face. Accessed: 2026-01-05}
}

@inproceedings{verl,
  title={Hybridflow: A flexible and efficient rlhf framework},
  author={Sheng, Guangming and Zhang, Chi and Ye, Zilingfeng and Wu, Xibin and Zhang, Wang and Zhang, Ru and Peng, Yanghua and Lin, Haibin and Wu, Chuan},
  booktitle={Proceedings of the Twentieth European Conference on Computer Systems},
  pages={1279--1297},
  year={2025}
}

@article{yan2025beyond,
  title={Beyond self-talk: A communication-centric survey of llm-based multi-agent systems},
  author={Yan, Bingyu and Zhou, Zhibo and Zhang, Litian and Zhang, Lian and Zhou, Ziyi and Miao, Dezhuang and Li, Zhoujun and Li, Chaozhuo and Zhang, Xiaoming},
  journal={arXiv preprint arXiv:2502.14321},
  year={2025}
}

@inproceedings{yan2026attack,
  title={Attack the messages, not the agents: A multi-round adaptive stealthy tampering framework for llm-mas},
  author={Yan, Bingyu and Zhang, Xiaoming and Zhou, Ziyi and Li, Chaozhuo and Zeng, Ruilin and Qi, Yirui and Wang, Tianbo and Zhang, Litian},
  booktitle={Proceedings of the AAAI Conference on Artificial Intelligence},
  volume={40},
  number={35},
  pages={29784--29792},
  year={2026}
}

@article{li2025loki,
  title={Loki's dance of illusions: A comprehensive survey of hallucination in large language models},
  author={Li, Chaozhuo and Wang, Pengbo and Wang, Chenxu and Zhang, Litian and Liu, Zheng and Ye, Qiwei and Xu, Yuanbo and Huang, Feiran and Zhang, Xi and Yu, Philip S},
  journal={arXiv preprint arXiv:2507.02870},
  year={2025}
}

@article{liu2026clawkeeper,
  title={ClawKeeper: Comprehensive Safety Protection for OpenClaw Agents Through Skills, Plugins, and Watchers},
  author={Liu, Songyang and Li, Chaozhuo and Wang, Chenxu and Hou, Jinyu and Chen, Zejian and Zhang, Litian and Liu, Zheng and Ye, Qiwei and Hei, Yiming and Zhang, Xi and others},
  journal={arXiv preprint arXiv:2603.24414},
  year={2026}
}

@article{wang2026devil,
  title={The devil behind moltbook: Anthropic safety is always vanishing in self-evolving AI societies},
  author={Wang, Chenxu and Li, Chaozhuo and Liu, Songyang and Chen, Zejian and Hou, Jinyu and Qi, Ji and Li, Rui and Zhang, Litian and Ye, Qiwei and Liu, Zheng and others},
  journal={arXiv preprint arXiv:2602.09877},
  year={2026}
}

@article{zhao2022learning,
  title={Learning on large-scale text-attributed graphs via variational inference},
  author={Zhao, Jianan and Qu, Meng and Li, Chaozhuo and Yan, Hao and Liu, Qian and Li, Rui and Xie, Xing and Tang, Jian},
  journal={arXiv preprint arXiv:2210.14709},
  year={2022}
}

@article{liu2025scales,
  title={The scales of justitia: A comprehensive survey on safety evaluation of llms},
  author={Liu, Songyang and Li, Chaozhuo and Qiu, Jiameng and Zhang, Xi and Huang, Feiran and Zhang, Litian and Hei, Yiming and Yu, Philip S},
  journal={arXiv preprint arXiv:2506.11094},
  year={2025}
}

@article{zhang2025llms,
  title={Llms are introvert},
  author={Zhang, Litian and Zhang, Xiaoming and Yan, Bingyu and Zhou, Ziyi and Zhang, Bo and Guan, Zhenyu and Zhang, Xi and Li, Chaozhuo},
  journal={arXiv preprint arXiv:2507.05638},
  year={2025}
}
